\documentclass[%
 reprint,
superscriptaddress,
nofootinbib,
 amsmath,amssymb,
 aps,
 showkeys,
]{revtex4-2}

\usepackage{graphicx}
\usepackage{dcolumn}
\usepackage{bm}
\usepackage{hyperref}
\usepackage[bottom]{footmisc}
\usepackage{footnote}
\usepackage{xcolor}
\hypersetup{breaklinks = true, colorlinks = true, citecolor = blue, linkcolor = blue, urlcolor = blue}

\usepackage[mathlines]{lineno}

\newcommand{\be}{\begin{equation}}
\newcommand{\ee}{\end{equation}}
\newcommand{\bea}{\begin{eqnarray}}
\newcommand{\eea}{\end{eqnarray}}

\newcommand{\AuAu}{Au$+$Au~}
\newcommand{\jpsi}{$J/\psi~$}
\newcommand{\psiTwoS}{$\psi(2s)$}
\newcommand{\ptSquare}{$p^{2}_{\mathrm{T}}~$}
\newcommand{\ptSquareNoSpace}{$p^{2}_{\mathrm{T}}$}

\newcommand{\wGammaN}{$W_{\mathrm{\gamma N}}~$}
\newcommand{\wGammaNnospace}{$W_{\mathrm{\gamma N}}$}
\newcommand{\sNNrhic}{$\sqrt{s_{_\mathrm{NN}}}=200$}

\begin{document}

\title{Exclusive $J/\psi$, $\psi(2s)$, and $e^{+}e^{-}$ pair production in Au$+$Au \\ultra-peripheral collisions at the Relativistic Heavy-Ion Collider}
\affiliation{Abilene Christian University, Abilene, Texas   79699}
\affiliation{AGH University of Krakow, FPACS, Cracow 30-059, Poland}
\affiliation{Argonne National Laboratory, Argonne, Illinois 60439}
\affiliation{American University in Cairo, New Cairo 11835, Egypt}
\affiliation{Ball State University, Muncie, Indiana, 47306}
\affiliation{Brookhaven National Laboratory, Upton, New York 11973}
\affiliation{University of Calabria \& INFN-Cosenza, Rende 87036, Italy}
\affiliation{University of California, Berkeley, California 94720}
\affiliation{University of California, Davis, California 95616}
\affiliation{University of California, Los Angeles, California 90095}
\affiliation{University of California, Riverside, California 92521}
\affiliation{Central China Normal University, Wuhan, Hubei 430079 }
\affiliation{University of Illinois at Chicago, Chicago, Illinois 60607}
\affiliation{Chongqing University, Chongqing, 401331}
\affiliation{Creighton University, Omaha, Nebraska 68178}
\affiliation{Czech Technical University in Prague, FNSPE, Prague 115 19, Czech Republic}
\affiliation{Technische Universit\"at Darmstadt, Darmstadt 64289, Germany}
\affiliation{National Institute of Technology Durgapur, Durgapur - 713209, India}
\affiliation{ELTE E\"otv\"os Lor\'and University, Budapest, Hungary H-1117}
\affiliation{Frankfurt Institute for Advanced Studies FIAS, Frankfurt 60438, Germany}
\affiliation{Fudan University, Shanghai, 200433 }
\affiliation{Guangxi Normal University, Guilin, 541004}
\affiliation{University of Heidelberg, Heidelberg 69120, Germany }
\affiliation{University of Houston, Houston, Texas 77204}
\affiliation{Huzhou University, Huzhou, Zhejiang  313000}
\affiliation{Indian Institute of Science Education and Research (IISER), Berhampur 760010 , India}
\affiliation{Indian Institute of Science Education and Research (IISER) Tirupati, Tirupati 517507, India}
\affiliation{Indian Institute Technology, Patna, Bihar 801106, India}
\affiliation{Indiana University, Bloomington, Indiana 47408}
\affiliation{Institute of Modern Physics, Chinese Academy of Sciences, Lanzhou, Gansu 730000 }
\affiliation{University of Jammu, Jammu 180001, India}
\affiliation{Kent State University, Kent, Ohio 44242}
\affiliation{University of Kentucky, Lexington, Kentucky 40506-0055}
\affiliation{Lawrence Berkeley National Laboratory, Berkeley, California 94720}
\affiliation{Lehigh University, Bethlehem, Pennsylvania 18015}
\affiliation{Max-Planck-Institut f\"ur Physik, Munich 80805, Germany}
\affiliation{Michigan State University, East Lansing, Michigan 48824}
\affiliation{National Institute of Science Education and Research, HBNI, Jatni 752050, India}
\affiliation{National Cheng Kung University, Tainan 70101 }
\affiliation{Nuclear Physics Institute of the CAS, Rez 250 68, Czech Republic}
\affiliation{The Ohio State University, Columbus, Ohio 43210}
\affiliation{Institute of Nuclear Physics PAN, Cracow 31-342, Poland}
\affiliation{Panjab University, Chandigarh 160014, India}
\affiliation{Purdue University, West Lafayette, Indiana 47907}
\affiliation{Rice University, Houston, Texas 77251}
\affiliation{Rutgers University, Piscataway, New Jersey 08854}
\affiliation{University of Science and Technology of China, Hefei, Anhui 230026}
\affiliation{South China Normal University, Guangzhou, Guangdong 510631}
\affiliation{Sejong University, Seoul, 05006, South Korea}
\affiliation{Shandong University, Qingdao, Shandong 266237}
\affiliation{Shanghai Institute of Applied Physics, Chinese Academy of Sciences, Shanghai 201800}
\affiliation{Southern Connecticut State University, New Haven, Connecticut 06515}
\affiliation{State University of New York, Stony Brook, New York 11794}
\affiliation{Instituto de Alta Investigaci\'on, Universidad de Tarapac\'a, Arica 1000000, Chile}
\affiliation{Temple University, Philadelphia, Pennsylvania 19122}
\affiliation{Texas A\&M University, College Station, Texas 77843}
\affiliation{University of Texas, Austin, Texas 78712}
\affiliation{Tsinghua University, Beijing 100084}
\affiliation{University of Tsukuba, Tsukuba, Ibaraki 305-8571, Japan}
\affiliation{University of Chinese Academy of Sciences, Beijing, 101408}
\affiliation{United States Naval Academy, Annapolis, Maryland 21402}
\affiliation{Valparaiso University, Valparaiso, Indiana 46383}
\affiliation{Variable Energy Cyclotron Centre, Kolkata 700064, India}
\affiliation{Warsaw University of Technology, Warsaw 00-661, Poland}
\affiliation{Wayne State University, Detroit, Michigan 48201}
\affiliation{Wuhan University of Science and Technology, Wuhan, Hubei 430065}
\affiliation{Yale University, New Haven, Connecticut 06520}

\author{M.~I.~Abdulhamid}\affiliation{American University in Cairo, New Cairo 11835, Egypt}
\author{B.~E.~Aboona}\affiliation{Texas A\&M University, College Station, Texas 77843}
\author{J.~Adam}\affiliation{Czech Technical University in Prague, FNSPE, Prague 115 19, Czech Republic}
\author{L.~Adamczyk}\affiliation{AGH University of Krakow, FPACS, Cracow 30-059, Poland}
\author{J.~R.~Adams}\affiliation{The Ohio State University, Columbus, Ohio 43210}
\author{I.~Aggarwal}\affiliation{Panjab University, Chandigarh 160014, India}
\author{M.~M.~Aggarwal}\affiliation{Panjab University, Chandigarh 160014, India}
\author{Z.~Ahammed}\affiliation{Variable Energy Cyclotron Centre, Kolkata 700064, India}
\author{E.~C.~Aschenauer}\affiliation{Brookhaven National Laboratory, Upton, New York 11973}
\author{S.~Aslam}\affiliation{Indian Institute Technology, Patna, Bihar 801106, India}
% ...

\collaboration{STAR Collaboration}\noaffiliation

\date{\today}
\begin{abstract} 
Measurements of exclusive $J/\psi$, $\psi(2s)$, and electron-positron ($e^{+}e^{-}$) pair photoproduction in \AuAu ultra-peripheral collisions are reported by the STAR experiment at \mbox{$\sqrt{s_{_\mathrm{NN}}}=200~\rm{GeV}$}. We report several first measurements at the Relativistic Heavy-Ion Collider, which are i) \jpsi photoproduction with large momentum transfer up to $2.2~\rm{[(GeV/c)^{2}]}$, ii) coherent \jpsi photoproduction associated with neutron emissions from nuclear breakup, iii) the rapidity dependence of incoherent \jpsi photoproduction, iv) the $\psi(2s)$ photoproduction cross section at mid-rapidity, and v) $e^{+}e^{-}$ pair photoproduction up to high invariant mass of 6 $\rm{GeV/c^2}$. For measurement ii), the coherent \jpsi total cross section of $\gamma  + \rm{Au} \rightarrow J/\psi + \rm{Au}$ as a function of the center-of-mass energy $W_{\rm{\gamma  N}}$ has been obtained without photon energy ambiguities. The data are quantitatively compared with the Monte Carlo models STARlight, Sartre, BeAGLE, and theoretical calculations of gluon saturation with color glass condensate, nuclear shadowing with leading twist approximation, Quantum Electrodynamics, and the Next-to-Leading Order perturbative QCD. At the photon-nucleon center-of-mass energy of 25.0 GeV, the coherent and incoherent \jpsi cross sections of Au nuclei are found to be $71\pm10\%$ and $36\pm7\%$, respectively, of that of free protons. These data provide an important experimental constraint for nuclear parton distribution functions and a unique opportunity to advance the understanding of the nuclear modification effect at the top RHIC energy.
\end{abstract}

\maketitle

\section{\label{sec:intro}Introduction}
In relativistic heavy-ion collisions, a large fraction of the total cross section is provided by photon-induced interactions, known as ultra-peripheral collisions (UPCs). Typically, UPCs take place when the impact parameter between two colliding nuclei is greater than the sum of their radii. The interaction is initiated by one or more photons emitted from the moving charged ions, where the photon interacts with the other nucleus. Due to the large mass of the heavy nucleus, the emitted photons have very small virtualities and transverse momenta~\cite{Cahn:1990jk,Klein:2016yzr}. UPCs are considered clean experimental probes to study cold Quantum Chromodynamics (QCD) in high energy nuclear collisions.

\begin{figure*}[thb]
\includegraphics[width=6in]{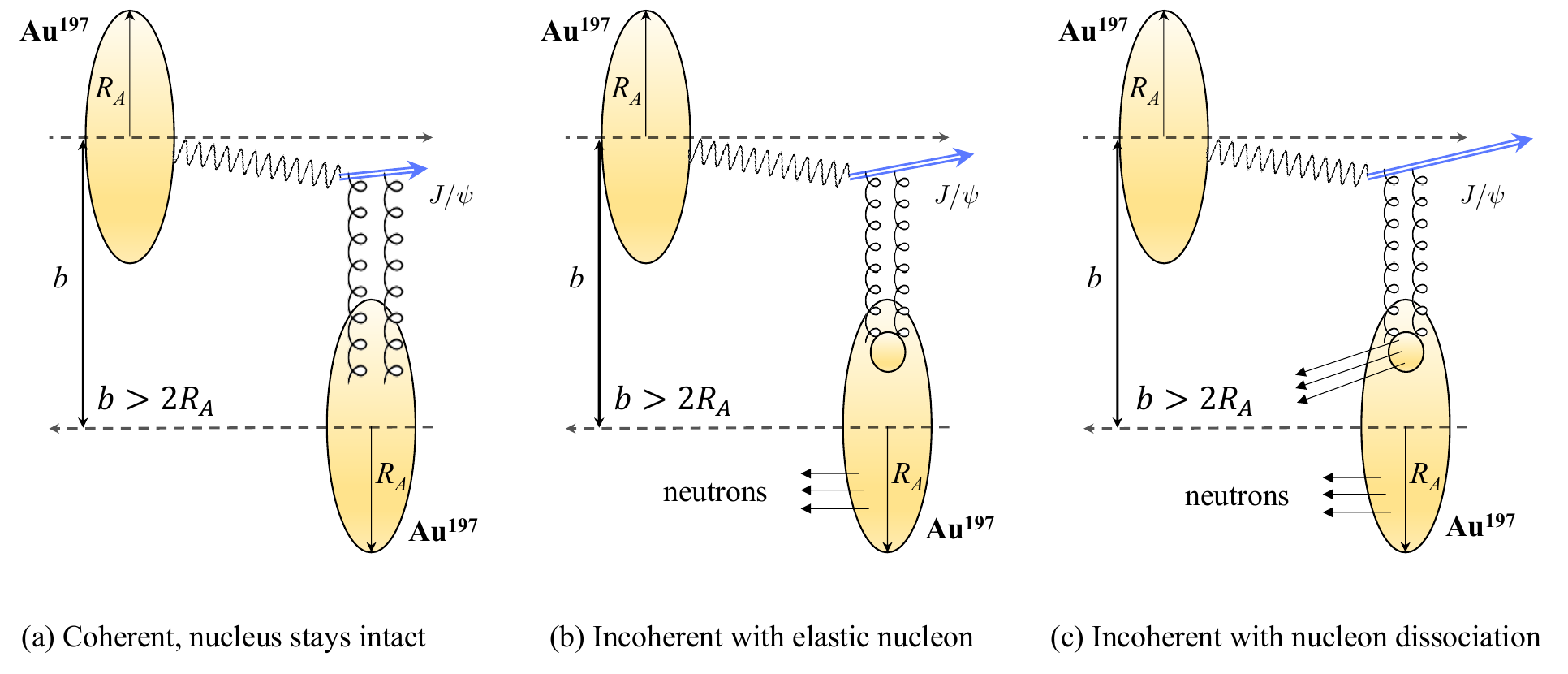}
  \caption{ \label{fig:intro:figure_0} Ultra-peripheral collisions at relativistic heavy ion colliders. (a) Coherent \jpsi photoproduction in Au$+$Au collisions where the nucleus stays intact; Coulomb excitation via soft photon exchange can break up the nucleus (not shown); (b) incoherent \jpsi photoproduction where the leading nucleon stays intact but the nucleus breaks up; (c) incoherent \jpsi photoproduction where the leading nucleon dissociates and the nucleus breaks up. }
\end{figure*}

Coherent diffractive Vector Meson (VM) photoproduction (nucleus stays intact),
through which the gluon density distribution of the nucleon and nucleus target can be directly probed,
has been extensively studied in recent years.
Photoproduction of \jpsi has been measured in heavy-ion UPCs with high precision
by the Large Hadron Collider (LHC)
experiments~\cite{Khachatryan:2016qhq,Abelev:2012ba,ALICE:2020ugp,ALICE:2021jnv,ALICE:2021tyx,ALICE:2021gpt,LHCb:2021hoq,CMS:2023snh,ALICE:2023jgu}. The resulting cross sections at low momentum fraction $x$ ($10^{-5}-10^{-3}$) were found to be significantly suppressed with respect to that of a free proton~\cite{ALICE:2012yye,Khachatryan:2016qhq,Abelev:2012ba,ALICE:2023jgu,CMS:2023snh}.
Calculations in the Leading Twist Approximation (LTA) strongly suggest that the suppression is caused by the nuclear shadowing effect~\cite{Strikman:2018mbu,Guzey:2013qza,Guzey:2018tlk}, while other models, e.g., the Color Dipole Model with gluon saturation and nucleon shape fluctuations~\cite{Sambasivam:2019gdd}, can also describe the UPC data qualitatively. The mechanism of gluon density modification in the nuclear environment at low-$x$ remains unknown. 
 
 Although UPCs provide experimental probes free of hadronic interactions, previous UPC measurements
 have an intrinsic ambiguity of the event kinematics. At each nonzero VM rapidity there are two possible photon energies, depending on which nucleus serves as the photon emitter. Therefore, the cross section measurement of VM photoproduction at any given nonzero rapidity includes contributions from a low and a high energy photon. The relative magnitude of the mixing depends solely on the photon flux at a given VM rapidity. Resolving this ambiguity would enable access to a wider phase space in kinematics. In order to understand the underlying physics mechanism of the modified parton density in nuclei, measurements with a wide range of kinematics from low-$x$ to high-$x$ are extremely important, since different physics models dominate at different kinematics. It has been suggested in Refs.~\cite{Guzey:2013jaa,Kryshen:2023bxy} that neutron emission from Coulomb excitation can be used to resolve the photon energy ambiguity. 
 
 Previous UPC measurements have mostly focused on coherent VM photoproduction, while the incoherent process has not been measured in detail (nucleus breaks up). However, the incoherent process has recently attracted increasing interest. Based on the Good-Walker paradigm~\cite{PhysRev.120.1857}, the incoherent VM cross section is sensitive to the event-by-event fluctuation of nuclear parton densities ~\cite{Mantysaari:2023jny}. Measurements of incoherent photoproduction have been proposed to investigate nuclear deformation, which is difficult to study in low energy nuclear experiments, as well as to study sub-nucleonic parton density fluctuations to understand the initial state condition of heavy-ion collisions~\cite{Arslandok:2023utm}.

At the Relativistic Heavy Ion Collider (RHIC) energy of $\sqrt{s_{_{\rm NN}}}$ = 200 GeV,
the kinematic phase space covered by the STAR experiment is complementary to that of the LHC. The per-nucleon center-of-mass energy, \wGammaNnospace\footnote{\wGammaN is 
 defined as $W_{\rm{\gamma N}}=\sqrt{2\left<E_{N}\right>M_{J/\psi}e^{-y}}$, where $E_{N}$ is the per-nucleon energy, $M_{J/\psi}$ and $y$ are the mass and rapidity of \jpsi particle.}, is 15--41 GeV within the \jpsi rapidity range $|y|<1.0$, which is similar to the previous STAR measurement of \jpsi photoproduction in the deuteron system~\cite{STAR:2021wwq}. The STAR kinematic region is at the transition ($x_{\rm parton} \approx 0.01$) between high-$x$ and low-$x$. In addition, the \jpsi momentum transfer $-t \simeq$ \ptSquare distribution can be measured at high \ptSquare with high precision, in a region expected to be sensitive not only to nucleon position fluctuations but also sub-nucleonic parton density event-by-event fluctuations.  
Different physics processes dominate in different regions of \ptSquareNoSpace. There are generally 3 types of processes, as illustrated in Fig.~\ref{fig:intro:figure_0}:
\begin{itemize}
\item \textbf{Coherent} $J/\psi$ production at low $p^{2}_{\rm{T}}$ (\mbox{$\lesssim 0.02~(\rm GeV/c)^{2}$}), where both nuclei stay intact; however, the nucleus can be broken up by additional soft photons via Coulomb excitation; the primary interaction is on the nucleus level.
\item \textbf{Incoherent elastic} $J/\psi$ production via elastic photon-nucleon scattering at intermediate $p^{2}_{\rm{T}}$ (\mbox{$\approx 0.02-0.5~(\rm [GeV/c)^{2}]$}), where the target nucleus may break up into fragments; the primary interaction is on the nucleon level.
\item \textbf{Incoherent dissociative} $J/\psi$ production with nucleon dissociation at high $p^{2}_{\rm{T}}$ ($ \gtrsim 0.5~\rm[(GeV/c)^{2}]$), where the leading nucleon (the nucleon undergoes a hard scattering) breaks up.
\end{itemize}

Note that the difference in Fig.~\ref{fig:intro:figure_0} (b) and (c) is not distinguishable event-by-event with the current detector setup in STAR. 

In this paper, along with results reported in a short Letter~\cite{STAR:2023nos}, we report measurements of both coherent and incoherent \jpsi photoproduction in Au$+$Au UPCs at $\sqrt{s_{_{\rm NN}}}$ = 200 GeV. The measurements are differential in momentum transfer \ptSquare and rapidity $y$, and performed for different neutron emission classes. Furthermore, we report the first measurement of $\psi(2s)$ photoproduction at RHIC
and measurements of the Quantum Electrodynamics (QED) process $\gamma\gamma\rightarrow e^{+}e^{-}$ in the invariant mass $m_{\rm{ee}}$ range of 2-6 $\rm GeV/c^{2}$. 

This paper is organized as follows: In Sec.~\ref{sec:models}, theoretical models that are quantitatively compared to the measured data are introduced. In Sec.~\ref{sec:detector}, a brief description of the STAR detector is given.
The data analysis is described in Sec.~\ref{sec:analysis}, including details of  signal extractions and cross sections and a summary of systematic uncertainties.
In Sec.~\ref{sec:results}, the main results are shown, followed by physics discussions and model validations in Sec.~\ref{sec:upcvalid}. Finally, a summary and outlook are discussed in Sec.~\ref{sec:conclusion}.

\section{\label{sec:models}Theoretical models}
Theoretical models provide important guidance for interpreting the data. In this paper, the data
have been compared quantitatively to several different models.
The models considered with brief descriptions are
as follows:

\begin{itemize}
    \item \textbf{STARlight.} A Monte Carlo event generator for simulating ultra-peripheral collisions in relativistic heavy-ion collisions~\cite{Klein:2016yzr}. It calculates the photon flux generated by heavy nuclei or protons via the equivalent photon approximation, requiring that there is no hadronic interaction, which is used in this analysis to derive the photon-nucleon cross section. The STARlight program can generate different UPC processes, e.g., VM photoproduction, QED two-photon ($\gamma\gamma$) processes, etc. The fundamental cross section for VM photoproduction, e.g., \jpsi in this analysis, is based on parameterized $\gamma p$ cross sections from HERA~\cite{H1:2005dtp,ZEUS:2002wfj}, nuclear form factors, and the photon flux. This model serves as a baseline for additional nuclear effects. In addition, the STARlight model, after tuning to describe the data, is used with the STAR detector simulation to correct detector effects in this analysis.

     \item \textbf{Sartre.} A Monte Carlo model for gluon saturation physics via exclusive VM photo- and electro-production in photon-nucleus collisions. The model applies nonlinear gluon evolution to calculate the scattering amplitude of a color-dipole, provided by the photon, and a target at a given impact parameter. Based on the Good-Walker paradigm, the model can predict both coherent and incoherent VM production, where the coherent process probes the average gluon density and the incoherent process is sensitive to the density fluctuation. Also known as the hot-spot model, the mechanism in incoherent VM production implements fluctuations of the parton and nucleon positions inside of a nucleon and a nucleus, respectively, their gluon densities, and their associated saturation scale $Q_{s}$. For this analysis, the comparisons to the measured data are based on Refs.~\cite{Toll:2012mb,Toll:2013gda}. This model is only valid for parton momentum fraction $x_{\rm{parton}}<0.01$, while it is compared with the STAR data at $x_{\rm{parton}}\approx 0.03$. The small mismatch to the data's kinematics needs to be taken into account when interpreting the data.

     \item \textbf{Color Glass Condensate (CGC)}. This is a theoretical calculation that has the same fundamental saturation physics mechanism as described in the Sartre model. Additionally, the CGC prediction for UPC \jpsi photoproduction has implemented the finite transverse momentum of the quasi-real photon and quantum interference effect when the UPC takes place in symmetric collision systems. For this analysis, the comparisons to the measured data are based on Ref.~\cite{Mantysaari:2022sux}. Similar to the Sartre model, the calculation is only valid for parton momentum fraction $x_{\rm{parton}}<0.01$~\cite{Mantysaari:2022sux}.
    
    \item \textbf{Nuclear Shadowing model with Leading Twist Approximation (LTA).} Nuclear shadowing model with Leading Twist Approximation (LTA) is a theoretical model based on Gribov-Glauber theory, the QCD factorization theorem, and HERA diffractive parton distribution functions (PDFs). For photoproduction of \jpsi in UPCs, the LTA predicts the cross section by dynamically modelling the multi-nucleon interaction at high energy. The case of no nuclear effects, the Impulse Approximation (IA), only considers a single nucleon interaction without final-state interactions. For this analysis, the comparisons to the measured data are based on Refs.~\cite{Guzey:2013jaa,Strikman:2005ze, Kryshen:2023bxy}. 

     \item \textbf{BeAGLE.} A general-purpose electron-nucleus ($e+$A) model, BeAGLE~\cite{Chang:2022hkt} is used for the description of incoherent photoproduction only. Details of this model are given in Refs.~\cite{Chang:2022hkt,Chang:2021jnu,Tu:2020ymk,Jentsch:2021qdp,PhysRevC.106.045202}. The comparison to heavy-ion (A$+$A) UPC cross sections is done by correcting the photon flux from $e+$A to A$+$A, where the $e+$A photon flux is based on PYTHIA 6~\cite{Sjostrand:2006za} and the A$+$A UPC photon flux is provided by the STARlight generator. BeAGLE uses the FLUKA program~\cite{Bohlen:2014buj,Ferrari:2005zk} to describe neutron emission from nuclei. For this analysis, the comparisons to the measured data are based on Ref.~\cite{Chang:2022hkt}.

    \item \textbf{QED.} Results for the purely QED process $\gamma \gamma \rightarrow e^+e^-$ are compared with a lowest-order QED calculation by Zha et.al~\cite{Zha:2018ywo,Zha:2018tlq}. The QED physics determining the photon flux follows the general principle of Weizsacker-Williams~\cite{vonWeizsacker:1934nji,Williams:1934ad},treating the electromagnetic fields in relativistic heavy-ion collisions as quasi-real photons.
    
    \item \textbf{Next-to-Leading Order (NLO) pQCD.} Based on the nuclear PDF EPPS21~\cite{Eskola:2021mjl}, the nuclear form factor, and the photon flux, the first NLO pQCD calculation for RHIC UPCs makes use of  parameters that were constrained by the LHC data. For this analysis, the comparisons to the measured data are based on Refs.~\cite{Eskola:2022vaf,Eskola:2022vpi}.
    
\end{itemize}

Note that each model has limitations and they are only compared to the data in applicable observables and kinematic regions.

\section{\label{sec:detector}Detector}
The Solenoidal Tracker At RHIC (STAR) detector~\cite{Ackermann:2002ad} and its subsystems have been thoroughly described in previous STAR papers~\cite{Adam:2018tdm,Adam:2020cwy}. This analysis utilizes several subsystems of the STAR detector. Charged particle tracking, including transverse momentum reconstruction and charge sign determination, is provided by the Time Projection Chamber (TPC)~\cite{Anderson:2003ur} positioned in a 0.5 Tesla longitudinal magnetic field. The TPC volume extends from 50 to 200 cm from the beam axis and covers pseudorapidities $|\eta|<1.0$ and the full azimuthal angle ($\phi$) range.
The TPC also provides particle energy loss information ($dE/dx$) used for particle identification. Surrounding the TPC is the Barrel Electromagnetic Calorimeter (BEMC)~\cite{Beddo:2002zx}, which is a
lead-scintillator sampling calorimeter approximately 20 radiation lengths in depth. The BEMC is segmented into 4800 optically isolated
towers covering the full azimuthal angle for pseudorapidities $|\eta|<1.0$. Between the TPC and BEMC is the Time Of Flight (TOF) system~\cite{Llope:2012zz}. It is finely segmented in $\eta$ and $\phi$ and provides timing signals for charged particles in the range $|\eta|<0.9$. There are two Beam-Beam Counters (BBCs)~\cite{Whitten:2008zz}, one on each side of the central STAR detector along the beam line, covering a pseudorapidity range of $3.4<|\eta|<5.0$. There are also two Zero Degree Calorimeters (ZDCs)~\cite{Ackermann:2002ad}, located $\pm18$ m from the center of STAR along the beam line, used to tag forward neutrons and monitor the luminosity.
The BEMC, TOF, BBCs and ZDCs provide fast signals which are used for triggering the STAR readout.

\section{\label{sec:analysis}Data analysis}
\subsection{\label{subsec:selection}Data selection}

The UPC data were collected by the STAR experiment during the 2016 \AuAu run, corresponding to an integrated luminosity of 13.5~$\rm nb^{-1}$ and approximately $24\times10^6$ UPC $J/\psi$-triggered events. The integrated luminosity is estimated based on the ZDC and VPD coincidence rate and the known hadronic Au$+$Au collision cross section. The final luminosity is corrected to the vertex $z$ range used in this analysis. 

\jpsi candidates are selected via the electron decay channel \jpsi$\rightarrow e^{+}e^{-}$. Based on this channel, the UPC \jpsi trigger is defined by BEMC energy deposits greater than $ \approx 0.7$ GeV in back-to-back azimuthal sextants of the BEMC. The TOF is required to have a hit multiplicity in the range of 2 to 6, and the BBCs are required to have no signal.
The BBC veto and upper limit on TOF multiplicity reject most hadronic Au+Au collisions.

Offline, pairs of tracks from a vertex within 100 cm of the center of STAR are considered. The tracks must extrapolate to energy deposits in the BEMC consistent with the trigger.
The tracks must have at least 15 points in the TPC to provide good momentum resolution,
and at least 11 $dE/dx$ measurements to provide good particle identification, out of a possible 45.
The measure $dE/dx$ information for a track was expressed as a
number of standard deviations from a particle identity hypothesis
$A$, $n\sigma_A$.
A measure of quality for the hypothesis for a pair is
$\chi^2_{AB} = (n\sigma_{A})^2 + (n\sigma_{B})^2$.
Tracks consistent with electron pairs were selected
by requiring $\chi^2_{ee}<10$
and those consistent with pion pairs were rejected
by requiring $\chi^2_{ee} < \chi^2_{\pi\pi}$.
Events with more than 6 significant energy deposits in the BEMC were rejected, providing further elimination of hadronic Au+Au collisions.
The back-to-back requirement in the trigger is inefficient for
low mass $m_{\rm{ee}}$, low $p_{\rm{T}}$ pairs,
thus only pairs with $m_{\rm{ee}} > 2$ GeV/c$^2$ are included in this analysis.
After applying all selection criteria, the sample includes approximately $3.9\times10^{4}$ pairs,
$7.9\times10^{3}$ of which are in the \jpsi mass range of $3.0<m_{\rm{ee}}<3.2$ GeV/c$^2$.

The selected pairs are predominantly opposite-sign $e^+e^-$,
containing the physics processes of interest.
A few percent of the pairs are like-sign, $e^+e^+$ or $e^-e^-$.
These are taken as an estimate of combinatorial background.
For all measurements, like-sign pairs are subtracted from
opposite-sign for final data distributions,
e.g. $m_{\rm{ee}}$ or $p_{\rm{T}}$.

\subsection{\label{subsec:simulation}Simulation}

In order to correct detector effects, the STARlight model that generates the \jpsi decay to two electrons and background contributions have been passed through STAR detector simulations. Specifically, the following processes with final state $e^+e^-$ are simulated:
\begin{itemize}
    \item Coherent \jpsi$\rightarrow e^{+}e^{-}$
    \item Incoherent \jpsi$\rightarrow e^{+}e^{-}$ with elastic nucleon
    \item Incoherent \jpsi$\rightarrow e^{+}e^{-}$ with nucleon dissociation
    \item Coherent $\psi(2s)\rightarrow e^{+}e^{-}$
    \item Coherent $\psi(2s) \rightarrow$\jpsi$+X$,
    followed by \jpsi$\rightarrow e^{+}e^{-}$ (feeddown)
    \item Two-photon interaction $\gamma\gamma\rightarrow e^{+}e^{-}$
\end{itemize}
As noted above, STARlight provides the basis of events used for simulation.
However, two of these processes are not included in the STARlight program.
The feeddown process $\psi(2s) \rightarrow$\jpsi$+X$ is modeled which uses the STARlight output from $\psi(2s)\rightarrow e^{+}e^{-}$ to define the $\psi(2s)$ momentum vector
and then generates the \jpsi$+X$ final state.
The incoherent \jpsi with nucleon dissociation is obtained by
reweighting the STARlight elastic incoherent \jpsi $p_{\rm{T}}$ distribution to the H1 nucleon dissociation
parameterization~\cite{Alexa:2013xxa}.

\begin{figure*}[thb]
\includegraphics[width=2.2in]{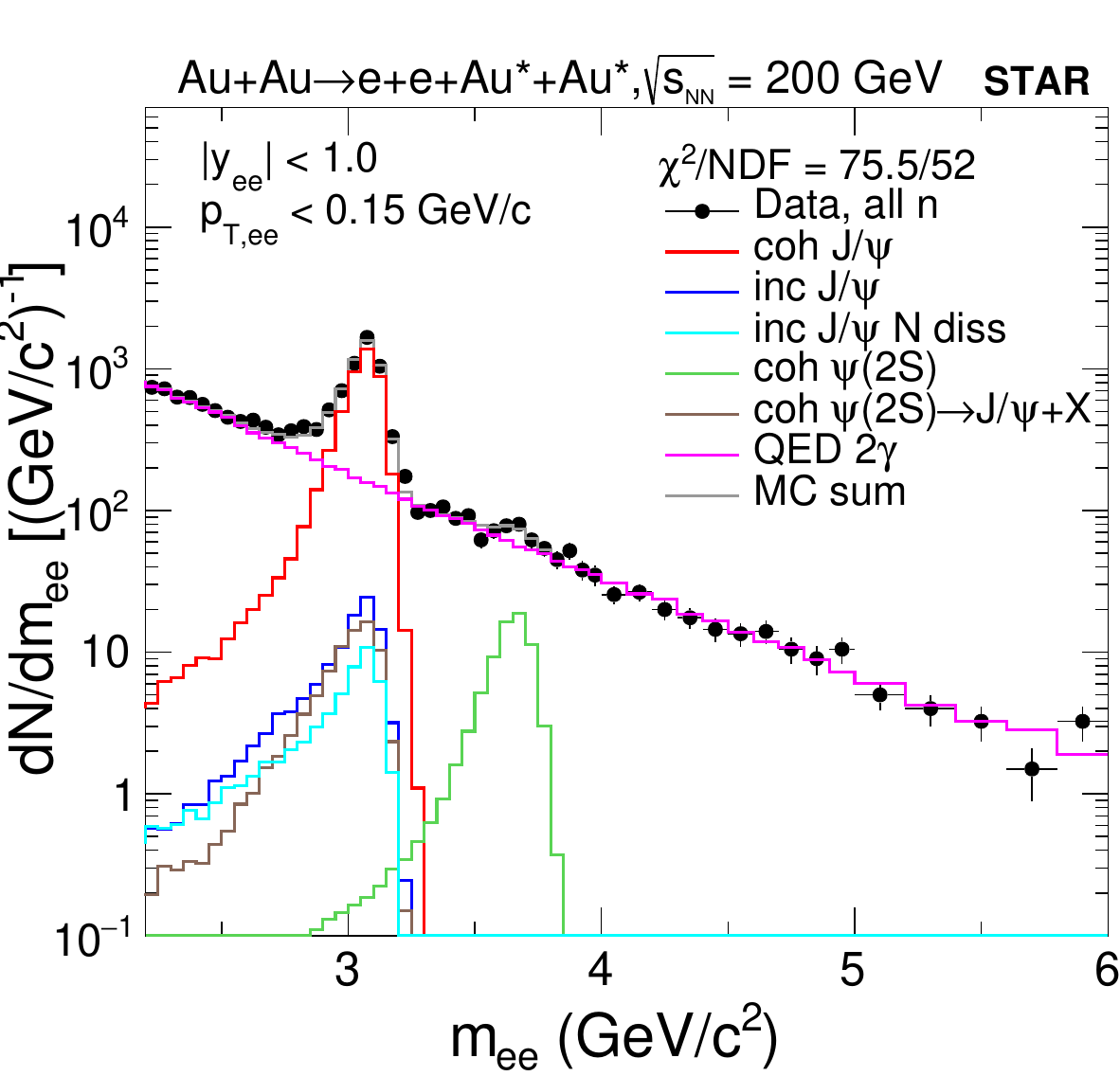}
\includegraphics[width=2.2in]{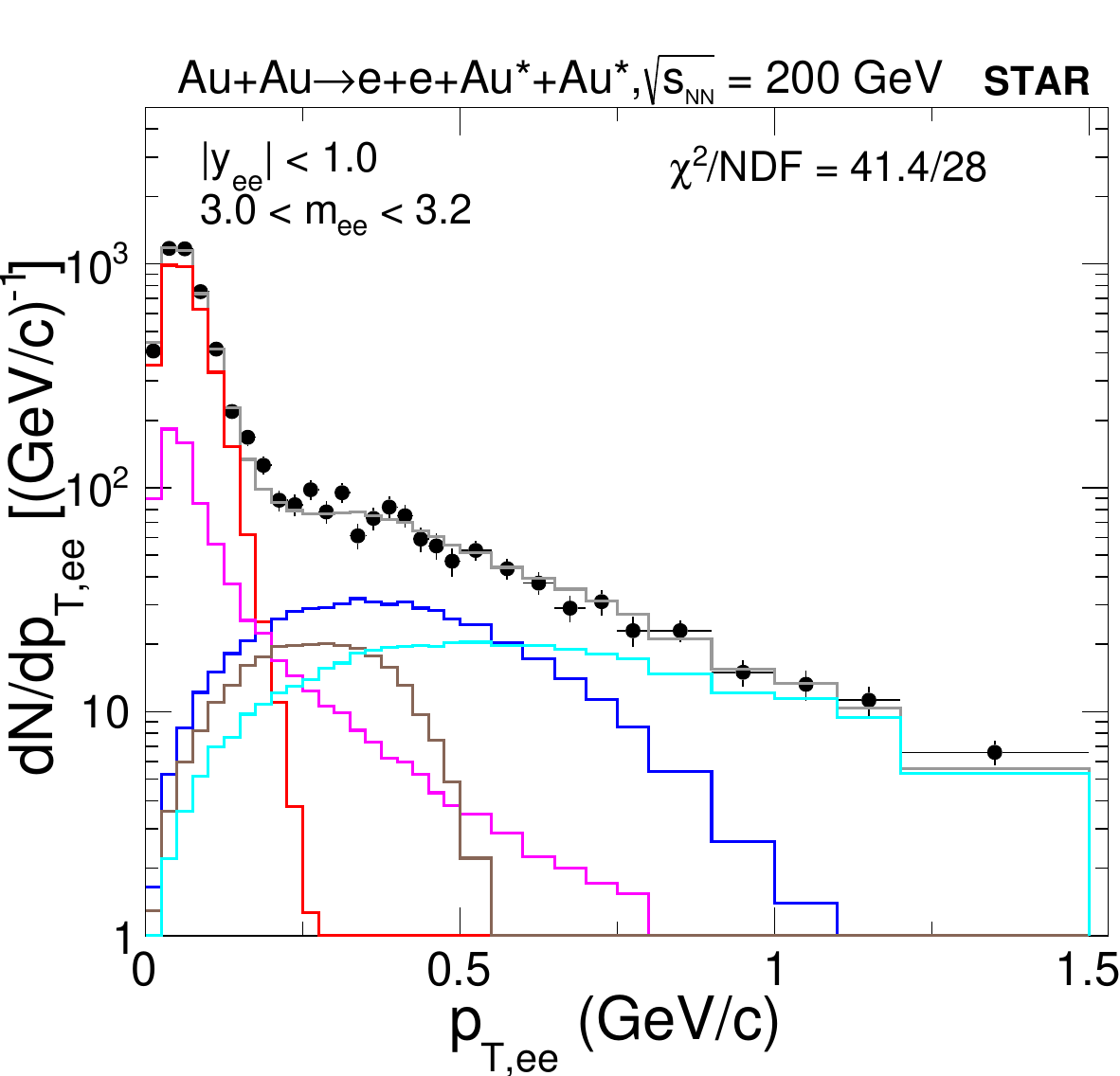}
\includegraphics[width=2.2in]{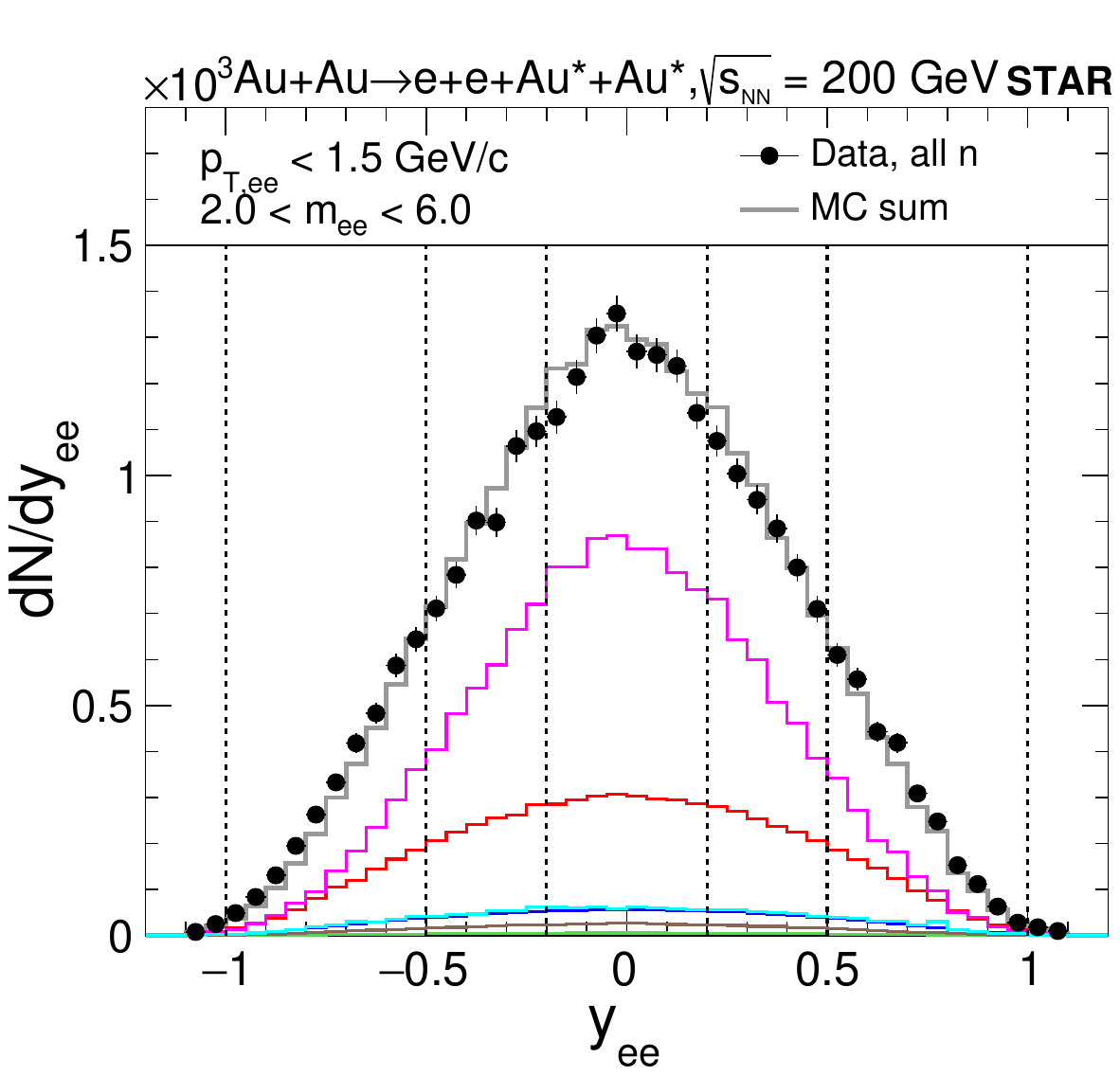}
  \caption{ \label{fig:res:figure_1} Invariant mass $m_{\rm{ee}}$, transverse momentum $p_{\rm T,ee}$, and rapidity $y_{\rm ee}$ of the electron pair candidates from \AuAu UPCs at $\sqrt{s_{_\mathrm{NN}}}=200$ GeV. They are shown in the left, middle, and right panel, respectively. Template fits from \jpsi coherent and incoherent production, QED processes, and $\psi(2s)$ are included. Only statistical uncertainties are shown as vertical bars.   }
\end{figure*}

Two improvements were made to the generated samples.
First, the STARlight $p_{\rm{T}}$ distributions of pairs from
coherent \jpsi and the two-photon processes
have higher $p_{\rm{T}}$ than observed in the data.
The STARlight events were reweighted by a factor of
$e^{- \Delta B p^{2}_{\rm T}}$ to describe the data,
with $\Delta B = 165$ [(GeV/c)$^{-2}]$ for coherent \jpsi and
$\Delta B = 260$ [(GeV/c)$^{-2}]$ for the two-photon process.
Second, the detector materials that can cause bremsstrahlung are not perfectly described by the STAR GEANT simulation, and STARlight does not
include radiative processes such as
\jpsi$\rightarrow e^{+}e^{-}\gamma$.
To account for this, a parallel sample of each process was
made, with one electron replaced with an electron plus a collinear photon
adding to the same energy.
The photons were generated with a
bremsstrahlung energy spectrum~\cite{PDG}.

The generated events were passed through the
simulation chain of the STAR detector.
Data from an zero-bias sample (triggered on colliding bunch crossing only) of
events recorded by STAR were added to the simulated
events to reproduce the underlying
activity in the TPC during RHIC operation.
The output of this was passed through the same
reconstruction algorithms as used for the data.

The reconstructed simulation events were selected using the same
track and vertex criteria as applied to the data.
The trigger energy efficiency was measured using the data
and applied to the simulated events as a weight.
Similarly, the efficiency of matching tracks to BEMC energy
deposits was determined using $e^+e^-$ pairs from a sample
of data based on TOF triggering and selection.
This efficiency was also applied by weighting the simulated events.
The selected and weighted events were used to create
template invariant mass and $p_{\rm{T}}$ of the pair $m_{\rm{ee}}$ and $p_{\rm{T}}$ distributions for each simulated process.

\subsection{\label{subsec:sigext}Signal extraction}

Figure~\ref{fig:res:figure_1} shows the pair mass $m_{\rm{ee}}$,
transverse momentum $p_{\rm{T,ee}}$, and rapidity $y_{\rm{ee}}$ distributions.
The mass distribution shown considers only pairs at low $p_{\rm{T}} <  0.15$ GeV/c, where the
coherent \jpsi and $\gamma\gamma\rightarrow e^+e^-$ processes dominate.
The $p_{\rm{T}}$ distribution is in the \jpsi mass range of $3.0<m_{\rm{ee}}<3.2$ GeV/c$^2$.
The rapidity distribution includes the full selected data sample and
shows the bins used for further analysis:
$|y_{\rm{ee}}|<0.2$, $0.2<|y_{\rm{ee}}|<0.5$, and $0.5<|y_{\rm{ee}}|<1$.

Shown in Fig.~\ref{fig:res:figure_1} are the
process templates from the simulation.
Their sum is fit to the data $m_{\rm{ee}}$ and $p_{\rm{T}}$ distributions
by $\chi^2$ minimization. The fit is performed simultaneously on the $m_{\rm{ee}}$ and $p_{\rm{T}}$ distributions with the same underlying parameters, which are the coherent \jpsi yield, relative contribution between coherent and incoherent \jpsi production, incoherent \jpsi with and without nucleon dissociation, $\psi(2s)$ decay, and QED $\gamma\gamma$ process. 
It determines the fraction of extra radiative processes;
the result is sufficient to account for the extra radiative effects and bremsstrahlung.
The sums of all processes are also shown in Fig.~\ref{fig:res:figure_1},
demonstrating a good description of the data.
The rapidity distribution, not used for the fitting, demonstrates the quality of the fit.

The fit templates are used to subtract backgrounds to
the physics processes of interest.
For \jpsi $p_{\rm{T}}$ distributions, the two-photon and
$\psi(2s)$ templates are subtracted from the data.
For two-photon $m_{\rm{ee}}$ distributions, templates
for all other processes are subtracted from the data.
The statistical uncertainty from the fit for each subtracted
template contributes to the systematic uncertainty.

The simulated distributions are also used to determine
acceptance corrections.
The corrections are applied bin-by-bin to the
$p_{\rm{T}}$ and $m_{\rm{ee}}$ distributions.
The efficiency of the TOF $\geq 2$ hits requirement in the
trigger was determined using a TOF-independent trigger, such that the TOF requirement is a complete subset of this trigger;
the losses due to the $\leq 6$ TOF hits and BBC vetoes were
measured using a sample of zero-bias events (triggered on colliding bunch crossing only).
These two factors, 20\% and 4\%, respectively, were applied as scale factors (1.25 and 1.04) to
the final cross sections.

$J/\psi$ production is measured as a doubly
differential cross section
$d^2 \sigma/dp^{2}_{\rm{T}} \, dy$.
The cross section for each $p^{2}_{\rm{T}},y$ bin $i$ is calculated as:
\begin{equation}
    \frac{d^2 \sigma}{dp_{\rm T}^2 \, dy}_i =
    \frac{N_{raw,i}}{\epsilon_{trig} \cdot corr_i \cdot L \cdot BR \cdot \Delta {p^{2}_{\rm{T}}}_i \cdot 2\Delta y_i}
\end{equation}
where:
\begin{itemize}
    \item $N_{raw,i}$ is the number of data events in bin $i$
    \item $\epsilon_{trig}$ is the scale factor correction for trigger efficiency
    \item $corr_{i}$ is the acceptance and efficiency correction for bin $i$
    \item $L$ is the total luminosity
    \item $BR = 5.97\%$ is the branching ratio for J/$\psi \rightarrow e^+e^-$\cite{10.1093/ptep/ptaa104}.
    \item $\Delta {p^{2}_{\rm{T}}}_i$ and $\Delta y_i$ are the widths of 
    $p^{2}_{\rm{T}},y$ bin $i$;
    the factor of 2 accounts for events with
    $y<0$ and $y>0$.
\end{itemize}

The QED two-photon process is measured as a differential cross section
$d \sigma/dm_{\rm{ee}}$.
The cross section for each $m_{\rm{ee}}$
bin $i$ is calculated as:
\begin{equation}
    \frac{d \sigma}{dm_{\rm{ee}}}_i =
    \frac{N_{raw,i}}{\epsilon_{trig} \cdot corr_i \cdot L \cdot \Delta m_{\rm{ee},i}}
\end{equation}
where $\Delta m_{ee,i}$ is the width of bin $i$.

The fit templates are also used to separate the
coherent and incoherent components of \jpsi 
production, as described in
Section~\ref{subsec:rapdist}.
The measured distributions
$d^2 \sigma/dp^{2}_{\rm{T}} \, dy$
are integrated over a range of $p^{2}_{\rm{T}}$, and the
templates fit to the differential cross section are used to extrapolate the unmeasured region at low $p_{\rm{T}}$ and subtract the
contribution from the other component.
This gives the differential cross sections
$d\sigma/dy$ for coherent and incoherent
\jpsi production. Note that this analysis does not separate incoherent contributions with and without dissociation. Only the total incoherent \jpsi production is reported.

\begin{figure*}[thb]
\includegraphics[width=3.0in]{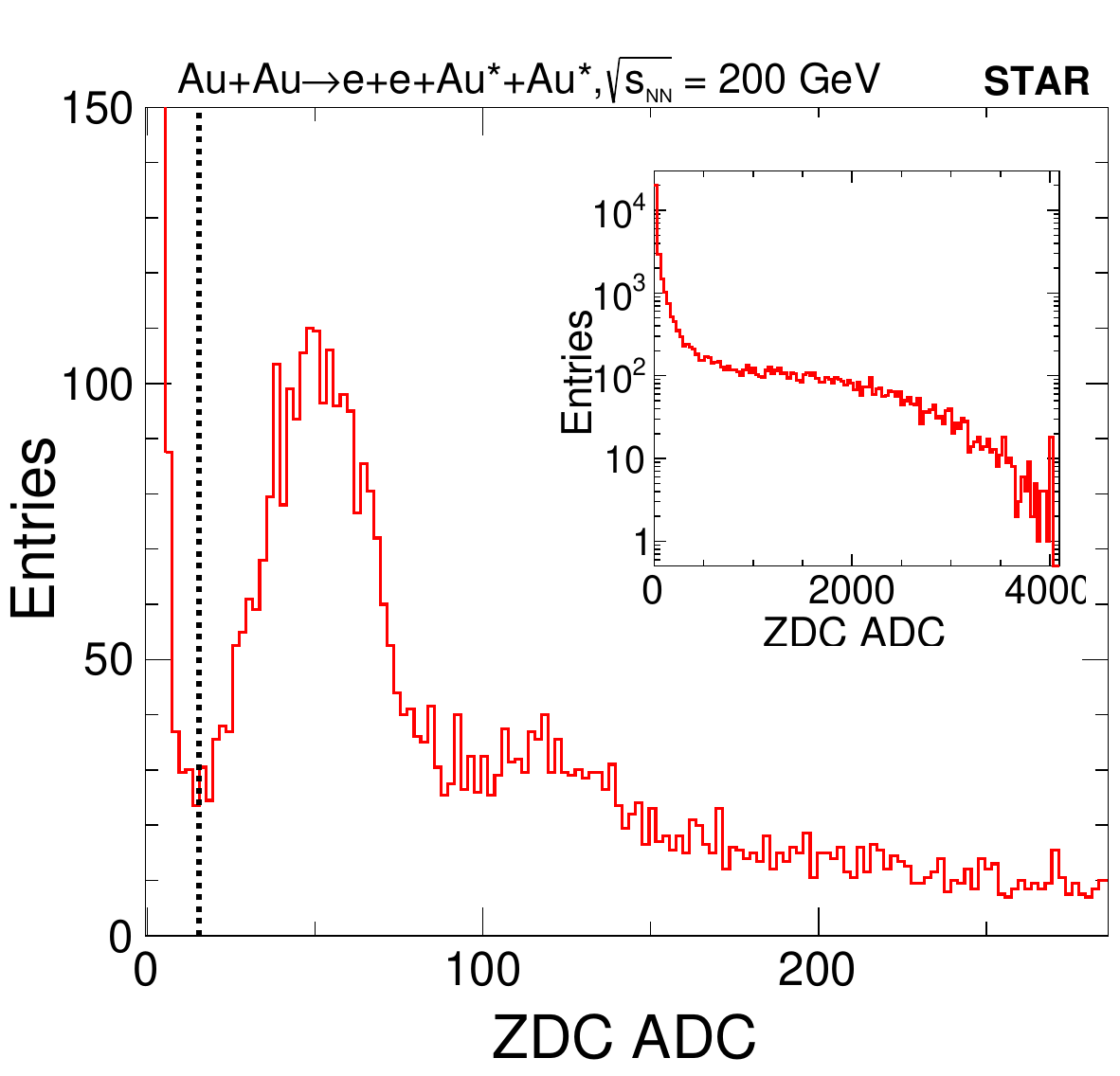}
\includegraphics[width=3.0in]{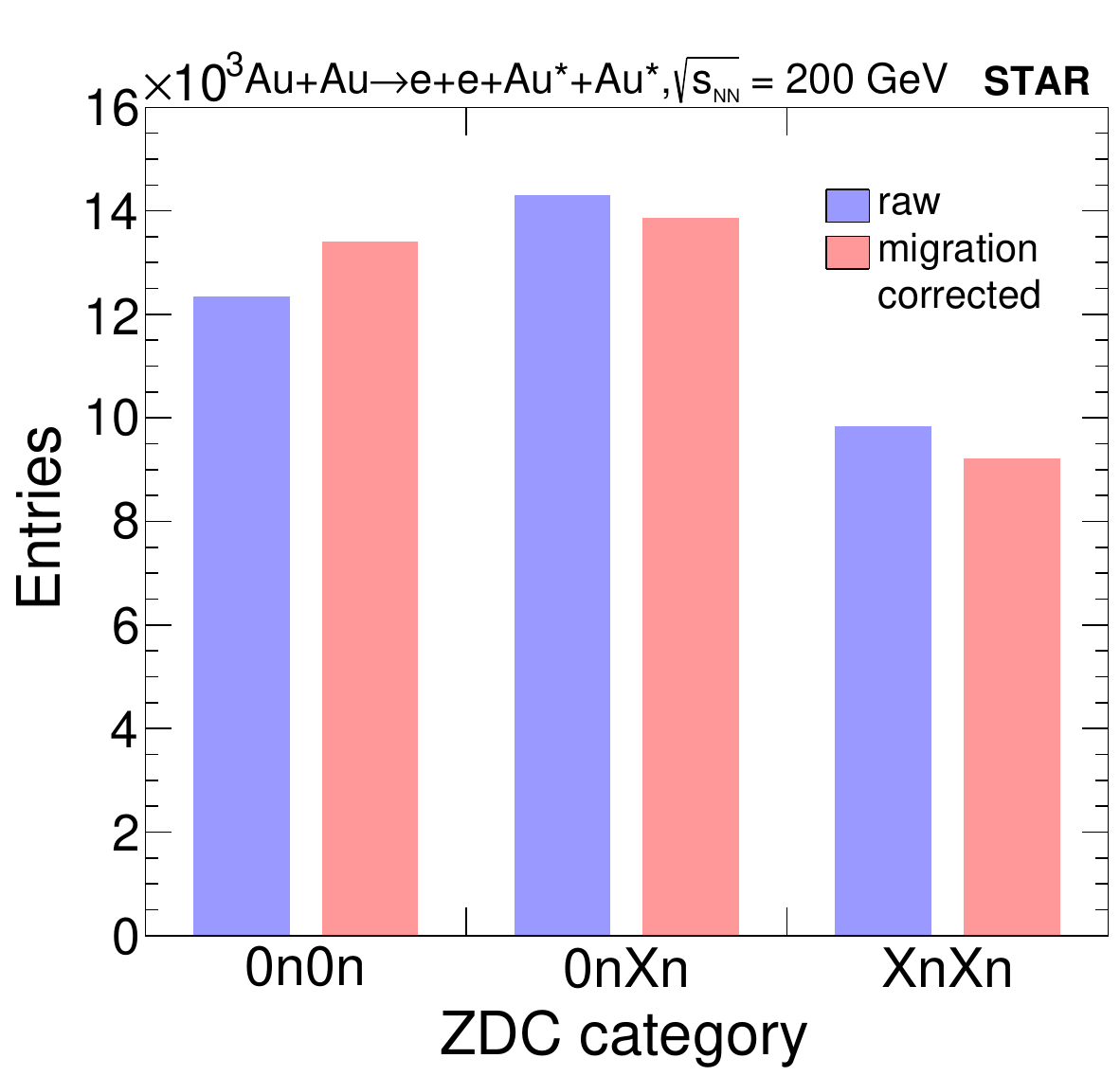}
  \caption{ \label{fig:res:figure_3} Left: the Analog-to-Digital Count (ADC) distribution from a Zero Degree Calorimeter (ZDC) in \AuAu UPCs at $\sqrt{s_{_\mathrm{NN}}}=200$ GeV. The separation between the noise peak near ADC = 0 and single neutron peak near ADC = 50 is clear. The inset shows the full ADC range with a cutoff corresponding to $\approx~$80 neutrons. Right: the ZDC category in terms of how many neutrons (0 or X) are shown before and after migration correction.  }
\end{figure*}

\subsection{\label{subsec:neuttag}Neutron tagging}

The left panel of Fig.~\ref{fig:res:figure_3} shows the pulse height
distribution from one of the ZDCs in the selected sample.
A clear single-neutron peak at $\rm ADC=50$ is evident, well separated
from the peak near zero; a two-neutron peak at $\rm ADC=120$ is also visible.
The inset shows the full ADC range, exhibiting an endpoint
at a pulse height corresponding to $\approx 80$ neutrons.
The distribution from the other ZDC is similar.
A neutron is defined as having a hit with a pulse height greater than 15 ADC counts in the ZDC, shown by
the dashed line in the figure. 

Neutron emissions in UPCs can provide insights to the VM production mechanism and impact parameter between the two nuclei beams. Specifically, they are categorized by their pattern of
neutron emission along the beamline,
which is measured by the ZDC hits on either side of
the central detector,
labeled as $0n$ (no hit) or $Xn$ ($\geq 1$ neutron) for each ZDC.
The categories are:
i) $0n0n$, neither ZDC hit;
ii) $0nXn$, one ZDC hit, one not hit;
iii) $XnXn$, both ZDCs hit.
The sum of i) to iii) is denoted as $all~n$.
The distribution of measured hit patterns for the full
data sample is shown by the uncorrected distribution in the right of
Fig.~\ref{fig:res:figure_3}.

Activity from other processes in the same RHIC bunch crossing
as the triggered event may include ZDC hits.
These will cause migrations to ZDC categories different from
those of the triggered event.
The migrations will be to higher ZDC multiplicity,
i.e. $0n0n$ to $0nXn$ and $XnXn$, $0nXn$ to $XnXn$.
Using a sample of zero-bias events, taken during the same time as the UPC $J/\psi$ trigger, the rate of overlaps was measured 
to be $f_1 = 3.8\%$ for a single hit in each of the ZDCs and not the other,
and $f_2 = 0.4\%$ for hits in both ZDCs;
the probability of no overlaps is then
$f_0 = 1 - 2 f_1 -f_2 = 92\%$.
This determines the migration between neutron categories;
the possibilities, with their probabilities in parentheses, are as follows:
\begin{itemize}
    \item
    A $0n0n$ event will remain a $0n0n$ event if there is
    no overlap ($f_0$),
    migrate to a $0nXn$ event if there is a single hit
    overlap in either ZDC ($2f_1$),
    or migrate to an $XnXn$ event if there is a double
    hit overlap ($f_2$).
    \item A $0nXn$ will remain a $0nXn$ event if there
    is no overlap or a single hit overlap in the same
    ZDC ($f_0+f_1$),
    or migrate to an $XnXn$ event if a single hit overlap
    is in the opposite ZDC or if there is a double hit
    overlap ($f_1+f_2$).
    \item An $XnXn$ event will always remain an $XnXn$
    event (1).
\end{itemize}
This can be written in matrix form, determining the
numbers of measured events $M_{\rm category}$
from the numbers of true events $N_{\rm category}$:
\begin{equation*}
    \left( 
    \begin{array}{c} M_{0n0n} \\ M_{0nXn} \\ M_{XnXn}
    \end{array} \right) =
    \left(
    \begin{array}{ccc}
    f_0 & 0 & 0 \\
    2f_1 & f_0+f_1 & 0 \\
    f_2 & f_1+f_2 & 1
    \end{array} \right) \cdot
    \left( 
    \begin{array}{c} N_{0n0n} \\ N_{0nXn}\\ N_{XnXn}
    \end{array} \right).
\end{equation*}
The matrix can be inverted to determine the numbers of
true categories from the measured ones.
As an example, the corrected distribution for the full 
data sample is shown in the right of Fig.~\ref{fig:res:figure_3}.
It shows the expected pattern,
with the number of uncorrected $0n0n$ events less than the corrected,
and the number of uncorrected $XnXn$ events is greater than the corrected.

Figure~\ref{fig:res:figure_3} demonstrates the migration correction
as applied to the total number of selected events.
For the $m_{\rm{ee}}$ and $p^{2}_{\rm{T}}$ distributions, the same procedure was
applied to each bin of the measured $0n0n$, $0nXn$, and $XnXn$
distributions, resulting in the corrected distributions for
the three neutron categories.

\subsection{\label{subsec:gamAuxsec}Coherent \jpsi cross sections in $\gamma$+Au collisions }

In heavy-ion UPCs, the hard photon may be
emitted by either beam nucleus, which would interact with the other nucleus. The cross section for coherent \jpsi photoproduction
as a function of rapidity $y$,
for a neutron emission pattern $ntag$,
is the sum of these two processes~\cite{Guzey:2013jaa}:
\begin{equation}
    \frac{d\sigma^{ntag}_{AuAu}}{dy} =
    \Phi^{ntag}_{T, \gamma}(k_+) \cdot \sigma_{\gamma Au}(W_+) +
    \Phi^{ntag}_{T, \gamma}(k_-) \cdot \sigma_{\gamma Au}(W_-)
    \label{eqn:dsigdy_dsigdw}
\end{equation}
Here $\Phi^{ntag}_{T, \gamma}(k)$ is the differential flux of hard photons based on STARlight~\cite{Klein:2016yzr}, 
$dN_{\gamma}/dy$ with energy
$k_{\pm} = 1/2 M_{J/\psi}e^{\mp y}$,
and $\sigma_{\gamma Au}(W)$ is the $\gamma$+Au cross section at
photon-nuclues center-of-mass energy
$W_{\pm}^2 = 4E_{N}k_{\pm}$. The $\pm$ of the energy corresponds to the ambiguity of the photon energy at a given rapidity due to the symmetric beam condition in Au$+$Au UPCs. Assuming the fundamental cross section of coherent \jpsi photoproduction is independent of neutron emissions from mutual Coulomb excitation and photon fluxes are dependent on the neutron emissions, the two unknown cross sections in this equation can be resolved.

For coherent \jpsi production, neutron emission occurs
through Coulomb excitation of the nuclei via
exchange of soft photons.
The different patterns of neutron emission result
from different soft photon exchanges at varying
impact parameters between the nuclei.
The different impact parameters determine different
hard photon fluxes for each neutron category,
$\Phi^{ntag}_{T, \gamma}$.
For the three neutron categories defined in this data set,
$ntag = $ $0n0n$, $0nXn$, and $XnXn$,
the three equations Eq.~\ref{eqn:dsigdy_dsigdw}
are independent.
A best fit determines the two
$\sigma_{\gamma Au}(W_{\pm})$ in terms of the
three measured $d\sigma^{ntag}_{AuAu}/dy$.

Note that this method only applies to coherent
$J/\psi$ photoproduction.
In incoherent \jpsi photoproduction,
mechanisms other than Coulomb excitation can
result in nuclear breakup and associated final
state neutrons.
This observed pattern of neutron emission
$ntag$ is no longer directly related to the
impact parameter constraints from Coulomb excitation, and does not define the hard photon
flux in Eq.~\ref{eqn:dsigdy_dsigdw}.

\begin{figure*}[thb]
\includegraphics[width=5.6in]{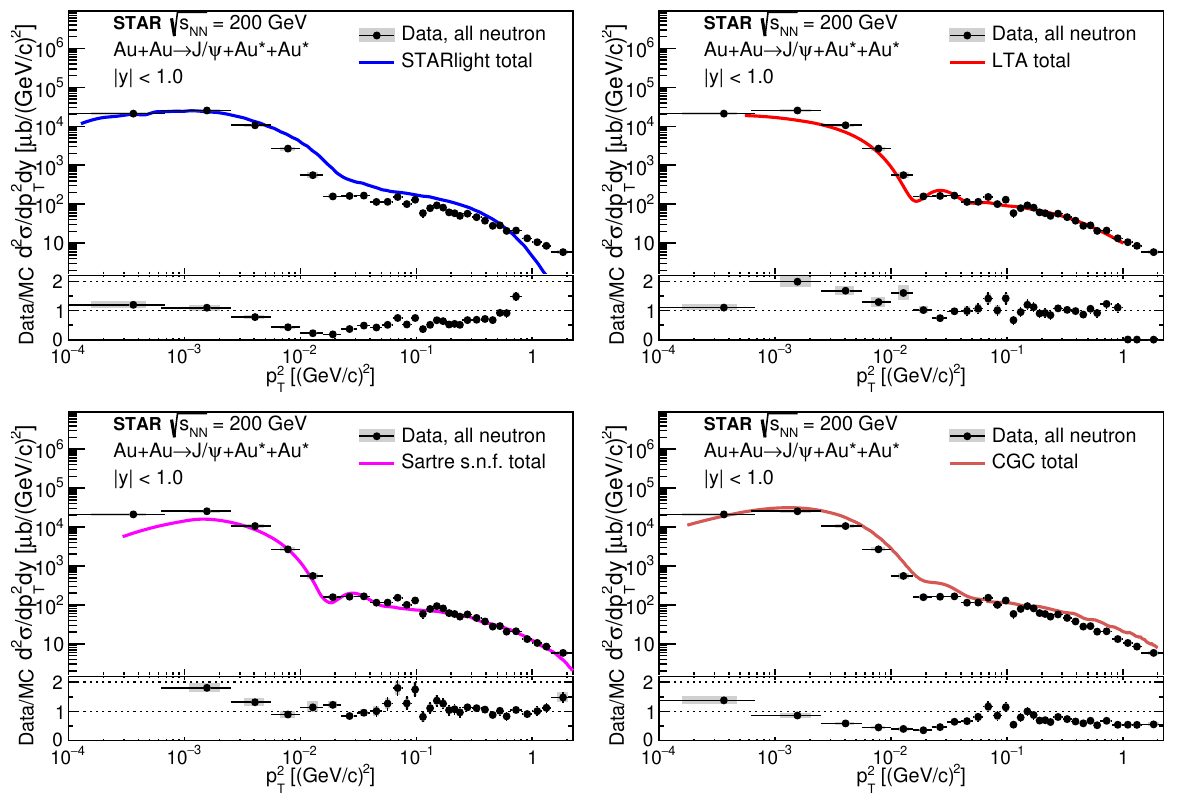}
  \caption{ \label{fig:res:figure_4a} Differential cross section $d^{2}\sigma/dp^{2}_{\mathrm{T}} dy$ of \jpsi photoproduction as a function of \ptSquare ($0.005 < p^{2}_{\rm{T}} < 2.2~\rm [(GeV/c)^{2}]$) in \AuAu UPCs at $\sqrt{s_{_{\rm{NN}}}}=200$ GeV. The rapidity of the \jpsi is $|y|<1.0$ and averaged \wGammaN is 25.0 GeV. MC model STARlight (upper left), nuclear shadowing of leading twist approximation calculation (upper right), Sartre MC prediction (lower left), and the color-glass-condense prediction (lower right) are compared with the data, presented as lines. Statistical uncertainty is represented by the error bars, and the systematic uncertainty is denoted as boxes. Ratio between data and models is shown in the lower sub-panel of each figure. There is a systematic uncertainty of 10\% from the integrated luminosity that is not shown.
   }
\end{figure*}

\subsection{\label{subsec:systuncert}Systematic uncertainties}

The systematic uncertainties on differential cross sections of $J/\psi$ and $\gamma\gamma\rightarrow e^{+}e^{-}$ are investigated separately.
Some systematic uncertainty sources are found to be \ptSquare or $m_{ee}$ dependent, which are presented bin-by-bin in the results shown later.
Other sources are applied as an overall scale.
Below, a brief description of each uncertainty source is discussed.  

Several factors contribute to the acceptance and efficiency corrections
for pair $m_{\rm{ee}}$ and $p_{\rm{T}}$ distributions.
The trigger energy efficiency was
modeled with an error function.
The uncertainty of trigger efficiency on
cross sections was
determined by varying the BEMC tower energy threshold by 0.11 GeV as the energy resolution of the BEMC, and results in cross section
uncertainties $\approx8\%$; it is greater at low $m_{\rm{ee}}$ close to the trigger threshold.
The efficiency of matching tracks to BEMC energy deposits
is evaluated with TOF matched electrons in the data, resulting in an uncertainty of $\approx5\%$
on pair reconstruction efficiency.
The uncertainty on weighting of STARlight to match $p_{\rm{T}}$ distributions is
only significant on the steeply falling coherent \jpsi
peak near $p_{\rm{T}} \approx 0.1$ GeV/c, where the pair uncertainty is up to 15\%.
The uncertainty from additional radiative events in the simulation
has an uncertainty of $\approx 2\%$ on the pair acceptance.

The background subtraction with fit templates described in
Section~\ref{subsec:sigext} has a statistical uncertainty from the fit.
It is largest in the region of $\psi(2s)$
feeddown, $0.2 < p_T < 0.4$ GeV/c, where 
it reaches $\approx$10\% for all neutron categories except for $0n0n$ where it dominates and reaches $\approx$50\%.
The uncertainty from background subtraction is $<2\%$ outside this $p_T$ region.
For the $\gamma\gamma$ $m_{ee}$ distributions
the J/$\psi$ region is not reported and the background subtraction outside this region is also negligible.

There is an overall 4\%
uncertainty on track and vertex reconstruction efficiency, which is found to be the same as in Ref.~\cite{STAR:2021wwq} as the data were taken in the same run period.

There is uncertainty on the \jpsi photoproduction cross section in $\gamma +$Au$\rightarrow J/\psi+$Au
from the photon flux used in extracting
this cross section.
It is estimated by varying the Au radius $\pm 0.5~\rm{fm}$, the same method as adopted in Ref.~\cite{STAR:2021wwq}.
The uncertainty on the cross section is found to be up to 3.5\%.

There is an uncertainty of 10\% on the luminosity measurement,
resulting in a scale uncertainty of 10\% on all cross sections,
which is not displayed with the data. This uncertainty is estimated by a special run - the Van-der-Meer scan~\cite{vanderMeer:1968zz}, which measures the instantaneous luminosity 
from the beam transverse sizes and currents; this sets the calibration factor used for the luminosity measurement. Note that there is an ongoing effort at RHIC trying to improve the uncertainty on the luminosity; however, it is not available at the time of this report and may be updated when it becomes available. 
All other uncertainties are shown with the displayed data points for all results presented hereafter.
All systematic uncertainty sources are added in quadrature to obtain the total systematic uncertainty, which has an
average value of $\approx$13.2\%. See Table~\ref{tab::sys} for a summary of uncertainty on the cross section measurements.

\begin{table}[tbh]
\caption{ \label{tab::sys} Summary of systematic uncertainty of $J/\psi$ and $e^{+}e^{-}$ photoproduction in Au$+$Au UPCs. Systematic uncertainty values are quoted for an average of each source, which are added in quadrature to be the total uncertainty. }
\fontsize{10}{15}\selectfont
\begin{tabular}{lc}
\hline
\hline
\textbf{Systematic sources}     & \textbf{Errors } \\ \hline
Trigger Efficiency (\%)         & 8     \\ 
BEMC Matching Efficiency (\%)   & 5     \\ 
Model template (\%)             & 2     \\ 
Electron Bremsstrahlung (\%)    & 2     \\ 
Signal Extraction (\%)          & 2     \\ 
Reconstruction efficiency (\%)  & 4     \\ 
Luminosity (\%)                 & 10    \\ \hline
\textbf{Total Uncertainty (\%)} & 14.7  \\ \hline
\hline
\end{tabular}
\end{table}

\section{\label{sec:results}Results}

\subsection{\label{subsec:pt2dist}
Momentum transfer distributions}

The square of the four-momentum transferred, $-t$,
between the incoming and outgoing nucleus
(transferred along the gluon lines in Fig.~\ref{fig:intro:figure_0})
characterizes the underlying nuclear geometry and its fluctuation in photon-nucleus interactions.
It has both longitudinal and transverse components:
\begin{align}
    t = t_{\parallel}+t_{\perp},\\
    -t_{\parallel} = M^2_{J/\psi}/(\gamma^2e^{\pm y}),\\
    -t_{\perp} = p^{2}_{\rm{T}}.
\end{align}
Here $\gamma$ is the Lorentz boost of the beam nuclei.
At the top RHIC energy, $t_{\parallel}$ is negligible due to the large $\gamma$ factor, thus,
\be
-t \approx p^{2}_{\rm{T}},
\ee
\noindent where $p_{\rm{T}}$ is the transverse momentum of the \jpsi.
The measured $p^{2}_{\rm{T}}$ of the \jpsi 
deviates slightly from $-t$ due to the
transverse momentum of the hard photon emitted
by the other nucleus, on average $\approx30$ MeV/c.
However, this smearing due to the photon $p_{\rm{T}}$ is relatively small and is included in some model calculations~\cite{Mantysaari:2022sux,Mantysaari:2023jny,Klein:2016yzr}, which can be directly compared to the measured  $p^{2}_{\rm{T}}$ distributions of the \jpsi in the data.

\begin{figure*}[thb]
\includegraphics[width=5.6in]{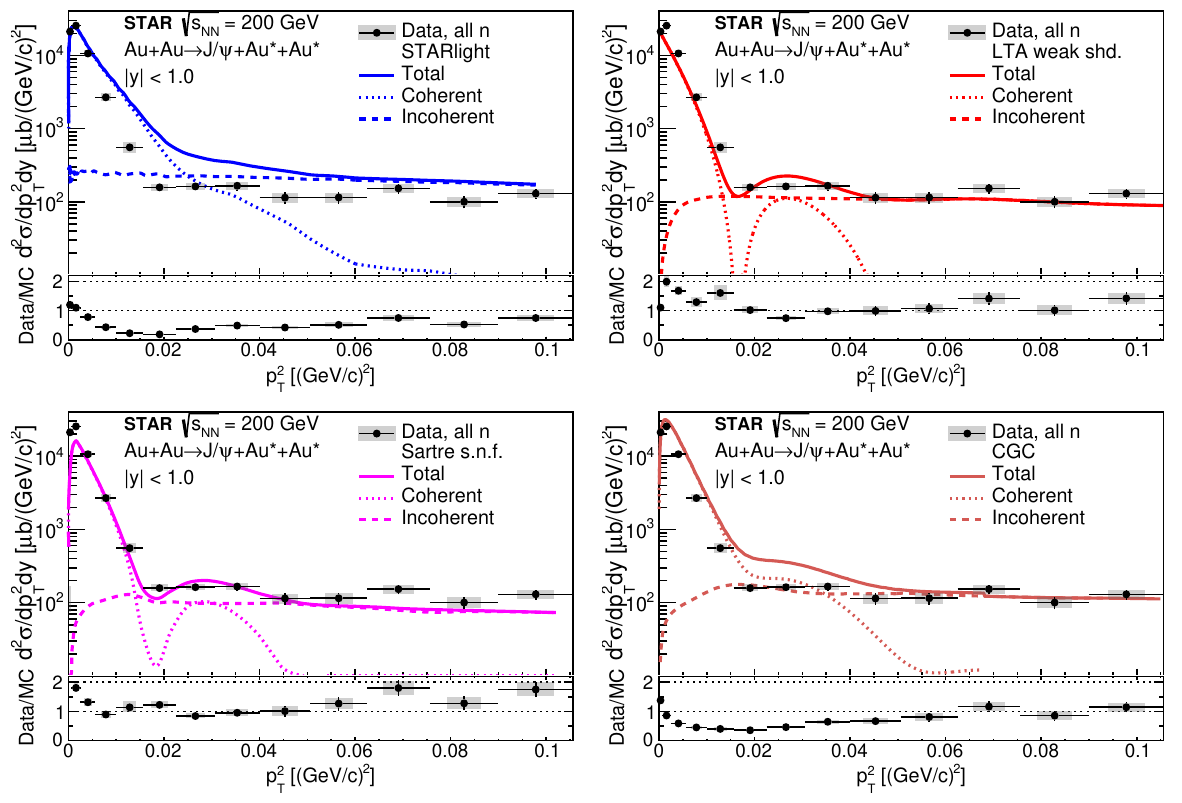}
  \caption{ \label{fig:res:figure_4b} Differential cross section $d^{2}\sigma/dp^{2}_{\mathrm{T}} dy$ of \jpsi photoproduction as a function of \ptSquare ($0 < p^{2}_{\rm{T}} < 0.12~\rm [(GeV/c)^{2}]$) in \AuAu UPCs at $\sqrt{s_{_{\rm{NN}}}}=200$ GeV. The rapidity of the \jpsi is $|y|<1.0$ and averaged \wGammaN is 25.0 GeV. MC model STARlight (upper left), nuclear shadowing of leading twist approximation calculation (upper right), Sartre MC prediction (lower left), and the color-glass-condensate prediction (lower right) are compared with the data, presented as lines. Statistical uncertainty is represented by the error bars, and the systematic uncertainty is denoted as boxes. The ratio between data and models is shown in the lower sub-panel of each figure. There is a systematic uncertainty of 10\% from the integrated luminosity that is not shown.
   }
\end{figure*}

In Fig.~\ref{fig:res:figure_4a} and Fig.~\ref{fig:res:figure_4b}, the \jpsi differential cross section, $d^2\sigma/dp^{2}_{\rm{T}}dy$ over the full measured range of \mbox{$p^{2}_{\rm{T}} < 2.2$ $\rm [(GeV/c)^{2}]$} and the low momentum transfer range \mbox{$p^{2}_{\rm{T}} < 0.12$ $\rm [(GeV/c)^{2}]$} is shown, respectively.
The data are identical among the four panels,
where each panel shows a comparison to a different theoretical model, indicated by the legend.
At lowest $p^{2}_{\rm{T}}$, the data exhibit a steep peak characteristic
of coherent \jpsi photoproduction.
As $p^{2}_{\rm{T}}$ increases the data follow a softly falling exponential,
indicative of incoherent photoproduction with scattering
off individual nucleons in the nucleus.
At higher $p^{2}_{\rm{T}}$, the distribution flattens in the
region where the scattered nucleon dissociates. The models compared with the data are the STARlight event generator, nuclear shadowing model LTA with weak shadowing mode, Sartre model with sub-nucleonic fluctuations, and CGC calculation with sub-nucleonic fluctuations. All models exhibit similar feature as the data, while the second peak or shoulder structure is caused by the coherent contribution. Furthermore, all models are presented with coherent, incoherent, and total cross sections in Fig.~\ref{fig:res:figure_4b}, while only the total cross sections are shown in Fig.~\ref{fig:res:figure_4a}. Ratios of data to the model total cross sections are shown at the bottom of each panel, where the model values are calculated by the integral of each bin instead of at the bin center.

Based on the data and model comparisons, the prediction of Sartre and LTA are found to be better in describing the magnitudes
and slopes of the coherent and incoherent components. 
The Sartre model has the same underlying physics model as the CGC framework with minor differences in the implementation. 
The CGC additionally accounts for the initial photon transverse momentum and interference effects. Both models consider the i) nonlinear gluon evolution in the target nucleus and ii) sub-nucleonic gluon density fluctuations.
The CGC calculation overpredicts the cross section by a factor of $\approx$1.5 at low ($\approx0.01~\rm{[(GeV/c)^{2}]}$) and high ($>1~\rm{[(GeV/c)^{2}]}$) \ptSquare   when compared to the data.
The same overprediction factor has been found in comparison to the LHC data~\cite{Mantysaari:2022sux}. Note that this prediction is explicitly made for the RHIC UPC measurement.

\begin{figure}[thb]
\includegraphics[width=3.4in]{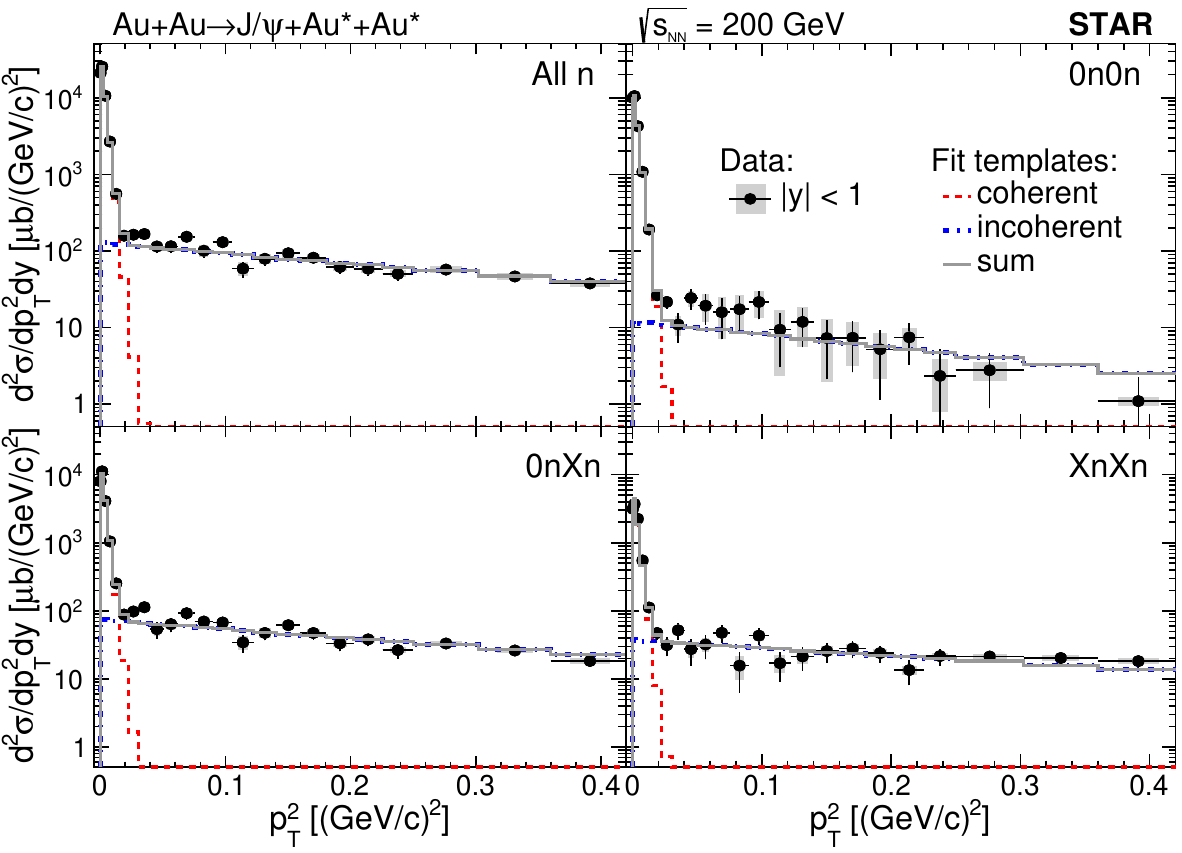}
  \caption{Differential cross section $d^{2}\sigma/dp^{2}_{\mathrm{T}} dy$ of \jpsi photoproduction as a function of \ptSquare in \AuAu UPCs at $\sqrt{s_{_{\rm{NN}}}}=200$ GeV. Four neutron emission classes, $all~n$, $0n0n$, $0nXn$, and $XnXn$ are shown in different panels. Coherent and incoherent template fits are shown as dashed lines, and the sum is shown as the solid line. Statistical uncertainty is represented by the error bars, and the systematic uncertainty is denoted as boxes. There is a systematic uncertainty of 10\% from the integrated luminosity that is not shown.  \label{fig:res:figure_7}  }
\end{figure}

Figure~\ref{fig:res:figure_7} shows the $p^{2}_{\rm{T}}$
distributions for all four neutron categories over
the full rapidity range $|y|<1$.
The distinction between the coherent peak and incoherent tail
is clearly visible due to the large difference in the slope,
suggesting that the total cross section for each process
can be measured with minimal extrapolation.
The template fits to the cross section for coherent and 
incoherent \jpsi production and their sum are also shown in
Fig.~\ref{fig:res:figure_7}.
The coherent cross section is determined
by integrating the data over \mbox{$p^{2}_{\rm{T}}<0.09$ $\rm [(GeV/c)^{2}]$}
and subtracting the incoherent template over the
same range.
The incoherent cross section is obtained
by integrating the data over \mbox{$p^{2}_{\rm{T}}>0.025$ $\rm [(GeV/c)^{2}]$}
and subtracting the coherent template over the
same range. To account for the low \ptSquare region, the cross section is scaled by the ratio of the
incoherent template integrated over all $p^{2}_{\rm{T}}$
to that of $p^{2}_{\rm{T}}>0.025$ $\rm [(GeV/c)^{2}]$.
These cross sections were measured for each of the three rapidity bins, resulting in the differential cross sections $d\sigma/dy$.

\subsection{\label{subsec:rapdist}Rapidity distributions}

\begin{figure}[thb]
\includegraphics[width=3.0in]{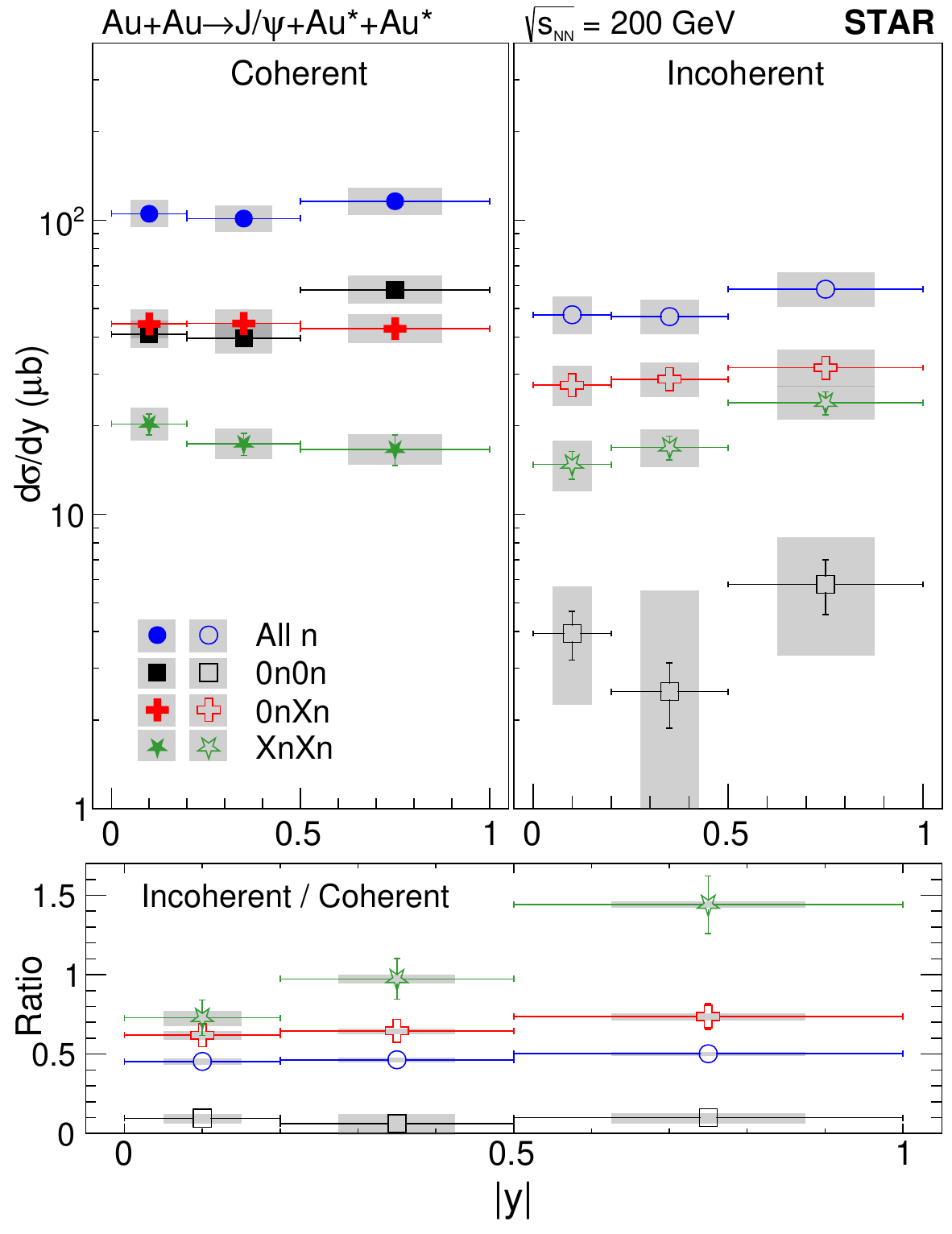}
  \caption{ \label{fig:res:figure_8_a} Differential cross section $d\sigma/dy$ for coherent, incoherent, and their ratio of \jpsi photoproduction as a function of $|y|$ in \AuAu UPCs at \sNNrhic~GeV. There is a systematic uncertainty of 10\% from the integrated luminosity that is not shown, while it is canceled in the ratio.}
\end{figure}

In Fig.~\ref{fig:res:figure_8_a}, the differential cross section, $d\sigma/dy$, of \jpsi photoproduction in the coherent (top left) and incoherent (top right) processes are presented.
The incoherent to coherent ratio is shown in the bottom panel. The measurements are also separated in different neutron emission configurations. 
The CGC calculation predicts that the ratio between the rapidity-dependent cross
sections for incoherent and coherent \jpsi photoproduction is
40\%, for nucleons that have event-by-event fluctuation in their gluon density~\cite{Mantysaari:2022sux}. On the other hand, with a static nucleon without the fluctuation, the prediction is around 20\%. The STAR data is found to be close to 40\%, which is shown in Fig.~\ref{fig:res:figure_8_a}, and favors the case of a fluctuating nucleon on this particular observable~\cite{Mantysaari:2022sux}. Although this is model dependent, the enhancement of incoherent cross section has been regarded as a result of the sub-nucleonic parton density fluctuation observed in the HERA data~\cite{Mantysaari:2016ykx}.

Although both coherent and incoherent photoproduction of \jpsi can be associated with neutron emissions, the underlying mechanisms are fundamentally different. For the coherent process, by definition, the target nucleus stays intact in \jpsi production (in the Good-Walker paradigm~\cite{PhysRev.120.1857}) ; however, additional soft photons can be emitted by either nucleus, which can excite one or both nuclei to break up and emit neutrons. Therefore, the differences among different neutron classes in coherent cross sections are purely due to the different probability of Coulomb excitation for a given impact parameter,
which is independent of the \jpsi production. On the other hand, neutron emission associated with incoherent \jpsi production is more directly related to the hard scattering. When the nucleus breaks up,
the fragments very often include neutrons.

\subsection{\label{subsec:gammaAuxsecdat} Coherent \jpsi cross section in $\gamma$+Au collisions}
\begin{figure}[thb]
\includegraphics[width=3.4in]{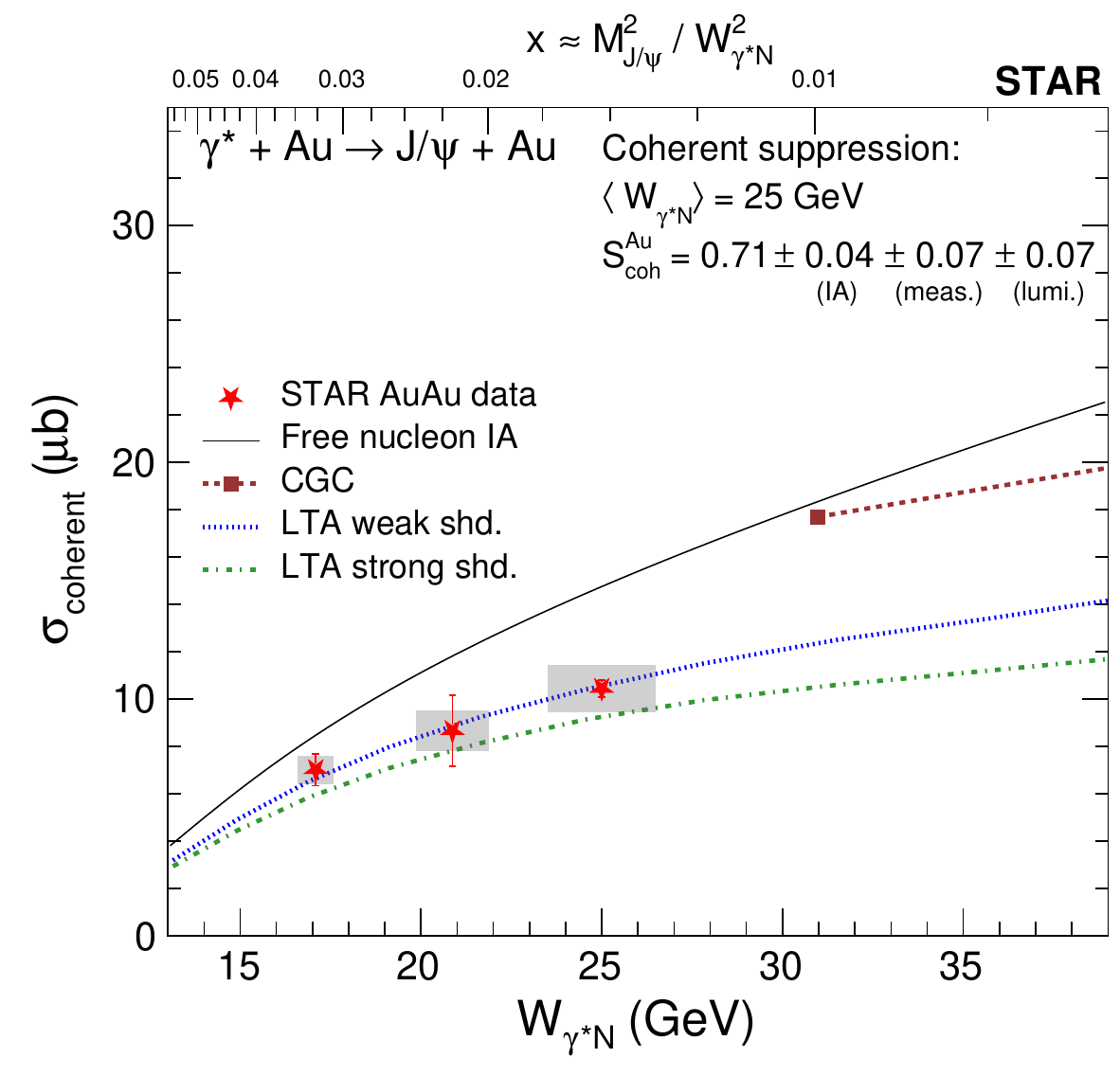}
  \caption{ \label{fig:res:figure_9} Total coherent \jpsi photoproduction cross section as a function of \wGammaN in \AuAu UPCs. The data are compared with an expectation of a free nucleon provided by the Impulse Approximation (IA)~\cite{Guzey:2013xba} and color glass condensate (CGC)~\cite{Mantysaari:2022sux}. The ratio between data and the Impulse Approximation at \wGammaN$=25.0$ GeV is the suppression factor, shown in the figure. Statistical uncertainty is represented by the error bars, and the systematic uncertainty is denoted as boxes. There is a systematic uncertainty of 10\% from the integrated luminosity that is not shown on the data points.}
\end{figure}
With the differential cross sections $d\sigma/dy$ for the three neutron categories $0n0n$, $0nXn$, and $XnXn$, the procedure to extract the $\gamma$+Au cross section described in Section~\ref{subsec:gamAuxsec} was followed. Each rapidity measurement determines two cross sections at different photon-nucleus center-of-mass energies $W_{\pm}$. For the present data the results for the higher energy $W_+$ are not statistically significant (consistent with zero) due to the low fraction of high energy flux compared to low energy. For the lowest rapidity interval the lower energy result at $W_-$ is also not significant.
Instead, the rapidity range $|y|<0.2$ was used to measure the
$\gamma$+Au cross section at the mean $W$ = 25.0 GeV. Note that there is a $-$0.5 GeV shift in the estimate of \wGammaN for rapidity $|y|<0.2$, caused by the higher photon flux of the lower energy photon contribution; however, the effect of this shift is found to be negligible.

In Fig.~\ref{fig:res:figure_9}, the total coherent \jpsi photoproduction as a function of \wGammaN is presented.
The data are free of photon energy ambiguity, based on the method discussed in Sec.~\ref{subsec:gamAuxsec}. The data are found to be suppressed with respect to the IA\cite{Guzey:2013xba} for all measured energies. Quantitatively, the suppression factor $\rm S^{Au}_{coh}$ is reported, which is the ratio between coherent \jpsi cross section and the IA.
It is found that $\rm S^{Au}_{coh}=0.71\pm0.04\pm0.07\pm0.07$ at \wGammaN$=25.0$ GeV. The first quoted error is the model uncertainty on IA~\cite{Guzey:2013xba} for Au nucleus and the second error is a combination of statistics and systematic uncertainties added in quadrature, while the third is from the scale
uncertainty of the integrated luminosity.
The reported STAR results in this analysis are the first measurements that contain no photon energy ambiguity at $y\neq0$ in a symmetric collision system at RHIC. The data can only be compared with the nuclear shadowing model, since the saturation-based models do not apply to \wGammaN lower than $\approx31$ GeV. It is found that the LTA with weak shadowing gives an excellent description of the data, while the LTA with strong shadowing predicts a slighly lower cross section. The data are found to be consistent with the LHC data at a similar energy range~\cite{ALICE:2023jgu, CMS:2023snh}. The implication of nuclear shadowing and its impact on the nuclear PDFs should be further investigated within its theoretical framework, and future p+A data with hard diffraction may shine new light on this question. Nevertheless,
the precise data presented here provide a stringent constraint on the nuclear PDFs and important information towards the understanding of the fundamental mechanism of such parton modification.

\subsection{Incoherent \jpsi cross section}
In Fig.~\ref{fig:res:figure_5_b}, the differential cross section of \jpsi production as a function of \ptSquare is shown for $|y| < 1.0$ for the $all~n$ neutron class. Due to the large rapidity range, the $\langle W_{\gamma N}\rangle$ is estimated to be 19.0 GeV based on the $all~n$ photon flux~\cite{Klein:2016yzr} and the energy-dependent cross section in $ep$ photoproduction~\cite{Alexa:2013xxa}. 

Starting above \ptSquare $> 0.02 \rm~[(GeV/c)^{2}]$, the cross section is dominated by the incoherent production. In order to compare with a free proton, the H1 published fit~\cite{Alexa:2013xxa} to $ep$ collisions is scaled down from $\langle W_{\gamma N}\rangle$ = 55 GeV to 19.0 GeV, based on the well measured energy-dependent cross sections parametrization as follows~\cite{Alexa:2013xxa}:
\begin{eqnarray}
    &\sigma_{el} = N_{el}\cdot (W_{\rm{\gamma N}}/90)^{\delta_{el}}, \\
    &\sigma_{pd} = N_{pd}\cdot (W_{\rm{\gamma N}}/90)^{\delta_{pd}}.
\end{eqnarray}

\noindent Here the $\sigma_{el}$ and $\sigma_{pd}$ are the proton elastic and proton dissociation cross section as a function of \wGammaNnospace. The parameters are $N_{el}=81\pm3 \rm{nb}$, $\delta_{el}=0.67\pm0.03$, $N_{pd}=66\pm7 \rm{nb}$, and $\delta_{pd}=0.42\pm0.05$. For the elastic proton case, UPC measurements in proton-lead UPCs at the LHC~\cite{ALICE:2018oyo} has a similar parametrization. Thus, the cross section ratios at different energies (e.g., 19.0 and 25.0 GeV) with respect to the H1 measured energy (55 GeV) are derived, which are used for obtaining the differential cross section $d\sigma/dt$ for proton elastic and proton dissociation at the STAR UPC kinematics.

For the differential cross section measurement, $d\sigma/dt$, as a function of momentum transfer $|t|$ at H1 can be fit by the following functions~\cite{Alexa:2013xxa}, 
\begin{eqnarray}
    & d\sigma_{el}/dt = N_{t,el}e^{-b_{el}|t|},\\
    & d\sigma_{pd}/dt = N_{t,pd}(1+(b_{pd}/n)|t|)^{-n}.
\end{eqnarray}

\noindent Here the parameter $N_{t,el}=213\pm18 ~\rm{nb/GeV^{2}}$, $b_{el}=4.3\pm0.2~\rm{GeV^{-2}}$, $N_{t,pd}=62\pm12~\rm{nb/GeV^{2}}$, $b_{pd}=1.6\pm0.2~\rm{GeV^{-2}}$, and $n$ is fixed at 3.58. 

Based on the above parametrization, the equivalent \AuAu UPC incoherent \jpsi cross section of a free proton can be rewritten as follows: 

\begin{figure}[thb]
\includegraphics[width=3.4in]{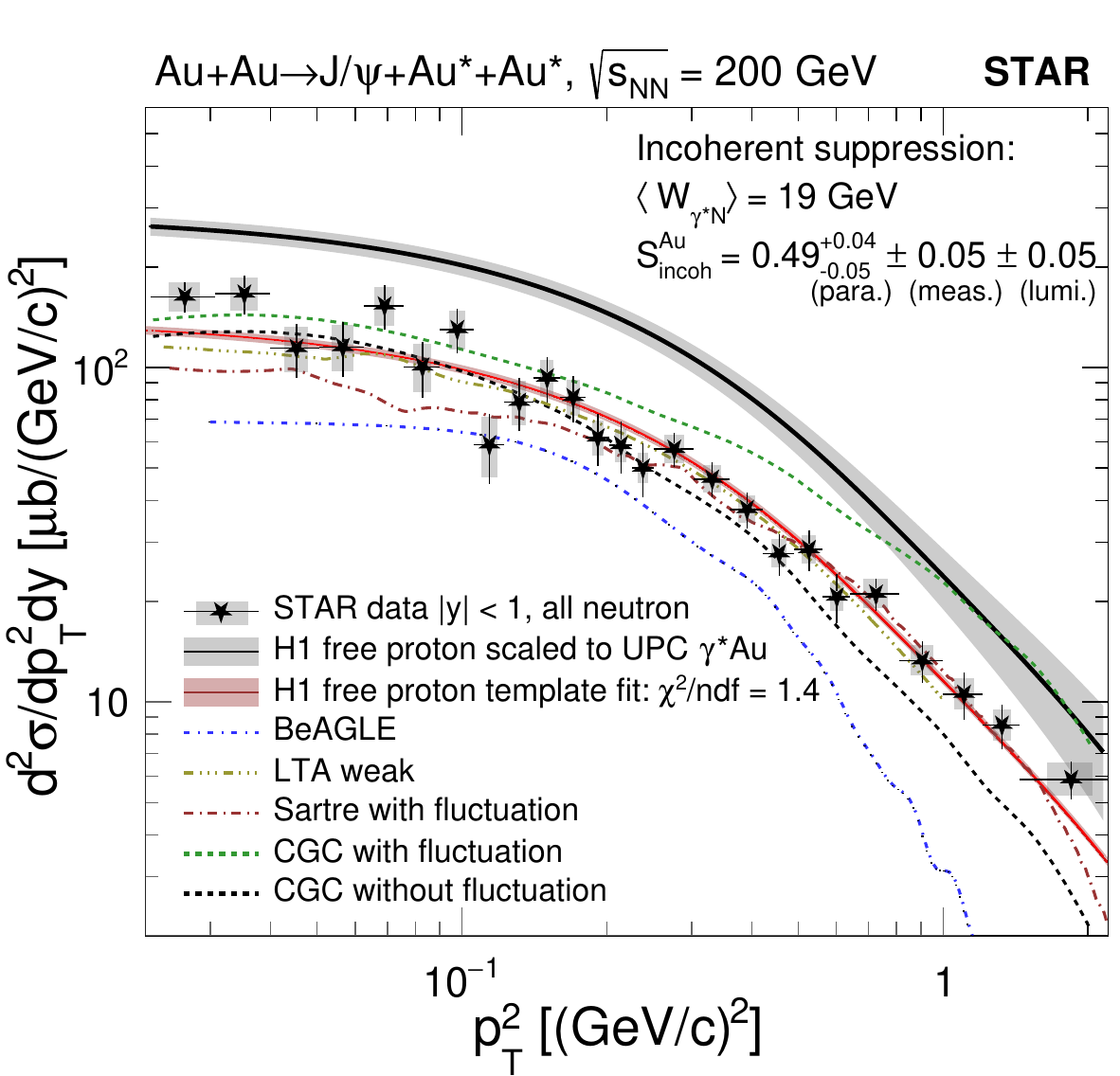}
  \caption{ \label{fig:res:figure_5_b} Incoherent \jpsi photoproduction differential cross section, $d^{2}\sigma/dp^{2}_{\mathrm{T}} dy$, as a function of \ptSquare is shown for $|y|<1.0$ without neutron class requirement. The H1 data in $ep$ collisions and its template fit to the STAR data are shown. The $1\sigma$ error of the fit is denoted as uncertainty bands. The ratio between the fit and the scaled H1 free data is the incoherent suppression factor, shown in the figure. The BeAGLE model~\cite{Chang:2022hkt}, the LTA weak shadowing calculation~\cite{Guzey:2013jaa}, Sartre model with sub-nucleonic fluctuation, and the CGC predictions~\cite{Mantysaari:2022sux}, are compared with the STAR data. Statistical uncertainty is represented by the error bars, and the systematic uncertainty is denoted as boxes. There is a systematic uncertainty of 10\% from the integrated luminosity that is not shown on the data points.   }
\end{figure}

\begin{eqnarray}
& &d^{2}\sigma_{\rm Au+Au\rightarrow J/\psi+Y}(\langle W_{\gamma N}\rangle)/dp^{2}_{\mathrm{T}}dy = \nonumber \\
& &2\Phi^{\rm all~n}_{T, \gamma}(\langle W_{\gamma p}\rangle)\cdot A \cdot  \left[ d\sigma_{\rm {\gamma +p\rightarrow J/\psi+Y}}(\langle W_{\gamma p}\rangle)/dt \right].
\end{eqnarray}

\noindent Here the $A$ is 197 for the Au nucleus, $ \Phi^{\rm all~n}_{T, \gamma}=5.03$ is the average transverse
photon flux at $y=0$ (the coefficient $2$ is the total flux from both beams being photon emitters), and $d\sigma_{\rm {\gamma+p\rightarrow J/\psi+Y}}(\langle W_{\gamma p}\rangle)/dt$ is the published H1 data at $\langle W_{\rm \gamma N}\rangle = \langle W_{\rm \gamma p}\rangle =19.0$ GeV scaled down from 55 GeV for both the elastic proton ($\rm{Y}=p$) and the proton dissociation ($\rm{Y}\ne p$). The notation is similar for the STAR data, where $\rm{Y}$ can be elastic nucleon or nucleon dissociation. Note that the published H1 data had been corrected for photon flux that is integrated over the phase space of $W_{\rm \gamma p}$, which is equivalent to the normalization of $1/dy$ in UPC measurements. The equivalent Au$+$Au UPC cross section for the free proton data is shown as the black solid line in Fig~\ref{fig:res:figure_5_b}, where the uncertainty band is propagated from the errors of the parametrization. 

Moreover, we use the H1 free proton data as a template to fit the STAR data with only the normalization constant as a free parameter. The integral of $d^{2}\sigma/dp^{2}_{\rm{T}}dy$ from \ptSquare = 0 to 2.2 $\rm{[(GeV/c)^{2}]}$ between the fit and the H1 data is defined as the incoherent suppression factor, $\rm S^{Au}_{incoh}$.
It is found that the $\rm S^{Au}_{incoh}$ is $0.49^{+0.04}_{-0.05}\pm0.05\pm0.05$ at \wGammaN$=19.0~\rm{GeV}$. For \wGammaN$=25.0~\rm{GeV}$ corresponding to the measurement within rapidity range $|y|<0.2$, the same procedure has been performed and the suppression factor is found to be $0.36^{+0.03}_{-0.04}\pm0.04\pm0.04$. Here the first uncertainty is the H1 parametrization uncertainty~\cite{Alexa:2013xxa}, the second one is from the measurement that includes statistical and systematic uncertainty, and the third is the scale uncertainty on the integrated luminosity. Therefore, the nuclear suppression in incoherent \jpsi photoproduction in \AuAu UPCs has been found to be stronger than that in the coherent case. This has been qualitatively predicted by the nuclear shadowing model LTA~\cite{Guzey:2013jaa, Kryshen:2023bxy}. 

Another observation is the similarity of shapes of the \ptSquare distributions between bound and free nucleons, which is quantified by the goodness-of-fit $\chi^{2}/ndf=1.4$. The 1 standard deviation (1$\sigma$) error is denoted by the uncertainty band. At very high \ptSquareNoSpace, there is a hint that the STAR data deviate above the H1 free proton template. However, measurements with higher precision and \ptSquare greater than 2.2 $\rm [(GeV/c)^{2}]$ are needed in order to draw conclusions. These data are the first quantitative measurement of incoherent \jpsi photoproduction of a bound nucleon in heavy nuclei.

Furthermore, the data are compared with different models. For the CGC calculations, 
the data are found to be in between the scenarios of strong sub-nucleonic parton density fluctuations and no fluctuations. It is not clear that the data directly supports either scenario. For the Sartre model, similar sub-nucleonic parton density fluctuations are included, which describes well the high \ptSquare tail but not the low \ptSquare behavior. Note that both the CGC and the Sartre model are calculated based on a higher energy configuration (corresponding to $x=0.01$) due to their model limitations. For the LTA with weak shadowing, the description of the data is very good. 
However, this is expected as the LTA model uses the HERA data parametrization. Finally, for the BeAGLE event generator, the cross section is underestimated for the entire \ptSquare range, which indicates using only the nuclear PDF, e.g. EPS09~\cite{Eskola:2009uj}, is not sufficient to describe the data.

\section{\label{sec:upcvalid}Physics discussions and Model validations}
\subsection{\label{subsec:breakup}Incoherent interactions and nuclear breakup}

\begin{figure}[thb]
\includegraphics[width=3.4in]{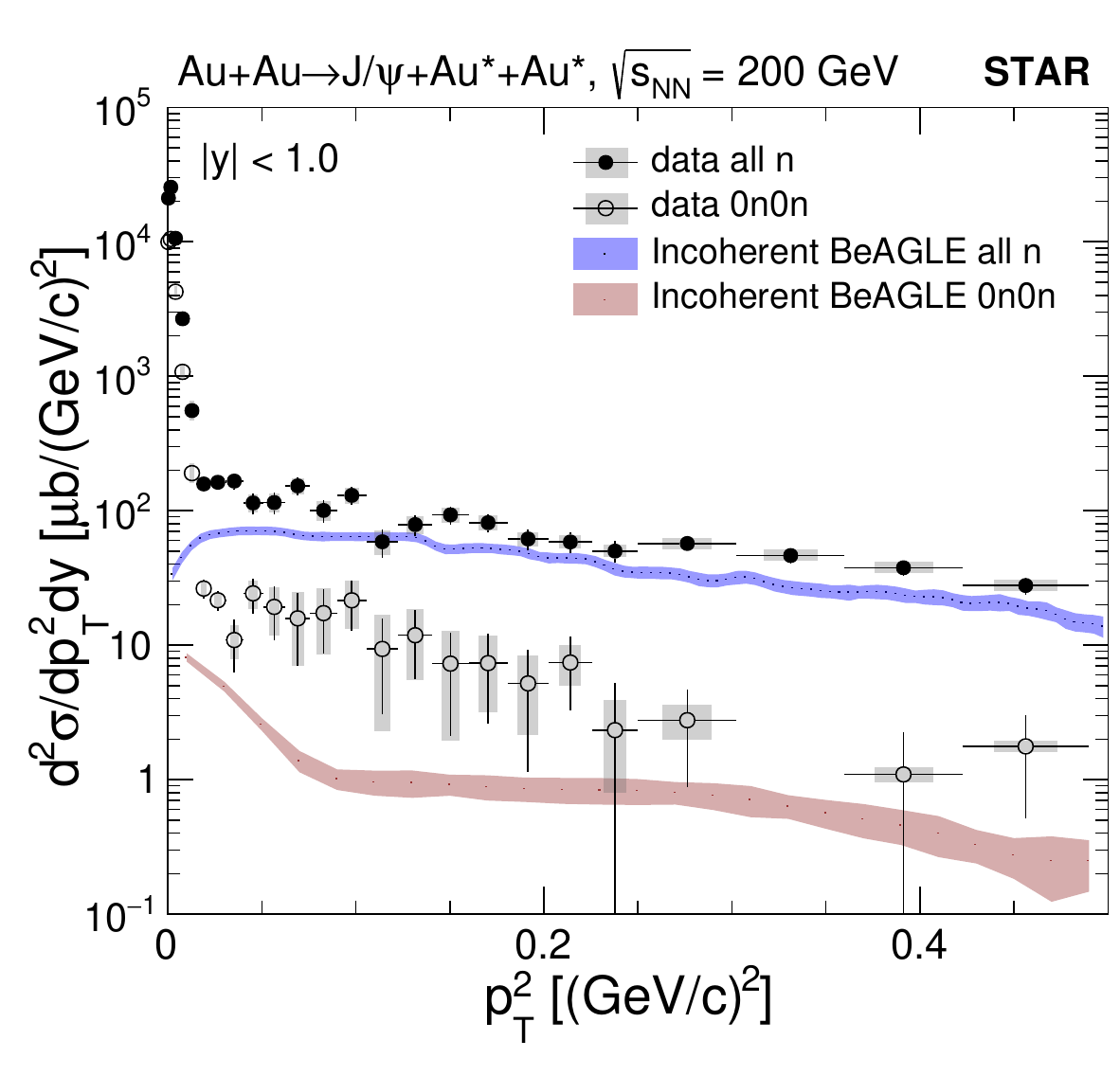}
  \caption{ \label{fig:res:figure_6} Differential cross section, $d^{2}\sigma/dp^{2}_{\mathrm{T}} dy$, of \jpsi photoproduction as a function of \ptSquare in \AuAu UPCs at \sNNrhic~GeV with all neutrons and $0n0n$ configuration. The data are compared with the incoherent BeAGLE simulation that has a STAR ZDC angular acceptance cut applied and the same neutron configurations.  Statistical uncertainty is represented by the error bars, and the systematic uncertainty is denoted as boxes. There is a systematic uncertainty of 10\% from the integrated luminosity that is not shown.  }
\end{figure}

Figure~\ref{fig:res:figure_6} shows the differential cross section of \jpsi photoproduction as a function of \ptSquare
in the full rapidity range $|y|<1$ for the $all~n$ 
case; the subset $0n0n$ neutron category is also shown.
In the coherent peak at lowest \ptSquareNoSpace,
the ratio of $0n0n$ to $all~n$ is $\approx40\%$,
consistent with the fraction of $0n0n$ photon flux,
where no neutrons are produced by
nuclear dissociation following
Coulomb excitation.
At higher \ptSquareNoSpace, where the incoherent processes dominates,
the additional mechanisms for neutron emission result
in the $0n0n$ fraction dropping to 10-20\%.

Also shown in Fig.~\ref{fig:res:figure_6} are the expectations
from the BeAGLE MC model, which only includes
incoherent photoproduction.
The full UPC photon flux is used for the $all~n$ case.
The corresponding photon flux is used for the $0n0n$ case to
describe no neutron emission through
nuclear dissociation following
Coulomb excitation.
The final state from BeAGLE is used to model the $0n0n$
state by requiring no breakup neutrons within the
ZDC acceptance of 2.5 mrad.
For the $all~n$ configuration, the BeAGLE model underestimates the cross section across the entire \ptSquare range, which is
possibly due to the poor description of the nuclear parton density (EPS09~\cite{Eskola:2009uj}) and lack of fluctuations.
Beyond this source, the difference between BeAGLE and data for $0n0n$ could be further related to how the evaporated neutrons are modeled in the nuclear breakup, which is one of the major concerns found in Ref.~\cite{Chang:2021jnu} for vetoing the incoherent production using the Far-forward detector system at the Electron-Ion Collider (EIC). This is the first measurement of \jpsi photoproduction associated with incoherent nuclear breakups, which is essential for improving $e\rm{A}$ MC models for the EIC.

\begin{figure}[thb]
\includegraphics[width=3.4in]{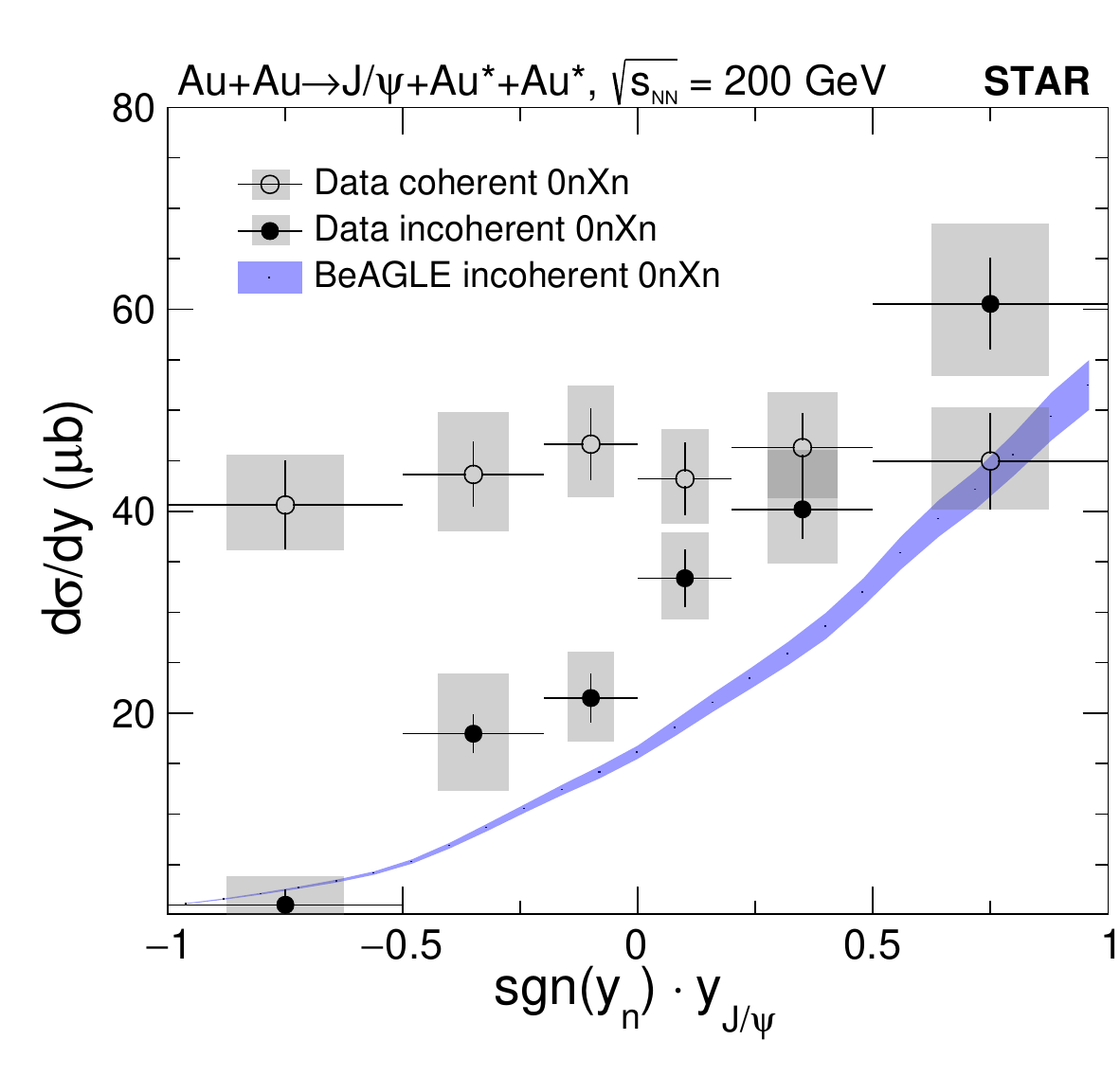}
  \caption{ \label{fig:res:figure_10} Coherent and incoherent differential cross section, $d\sigma/dy$, as a function of $y$ of \jpsi photoproduction with $0nXn$ neutron configuration in \AuAu UPCs at \sNNrhic~GeV. Here the negative $y$ direction has zero neutron ($0n$) and the positive $y$ direction has at least one neutron ($Xn$). The BeAGLE model is compared with the data. Statistical uncertainty is represented by the error bars, and the systematic uncertainty is denoted as boxes. There is a systematic uncertainty of 10\% from the integrated luminosity that is not shown. }
\end{figure}

The rapidity distributions for coherent and incoherent \jpsi
photoproduction are shown for the $0nXn$ neutron category
in Fig.~\ref{fig:res:figure_10}.
The asymmetric neutron configuration breaks the symmetry
of the collision. This allows a choice of sign for the \jpsi rapidity; here $y>0$ is chosen as the same direction as the
ZDC with a neutron hit in the $0nXn$ configuration.
The observed coherent \jpsi rapidity distribution
is symmetric under the transformation $y\rightarrow -y$. Neutron emission for coherent UPC photoproduction occurs only through Coulomb excitation, in which the neutron may be emitted
by either the target nucleus or the nucleus emitting the hard photon. Therefore, the neutron direction is not expected to be
correlated to the \jpsi direction, which is consistent with observation.

By contrast, in the incoherent photoproduction the target nucleus
usually breaks up in the hard interaction.
Resulting neutrons hitting a ZDC identify the direction
of the target nucleus.
For the $0nXn$ configuration this direction is unambiguous.

This is confirmed by the highly asymmetric rapidity distribution
for incoherent \jpsi photoproduction
as shown in Fig.~\ref{fig:res:figure_10}.
The figure also shows the result from a BeAGLE modelling of this reaction.
The model provides a good description of the shape of the data.
The difference in normalization may be due to the shortcomings
of BeAGLE cited earlier.
This asymmetric shape was also described in an earlier
discussion of the LTA model~\cite{Guzey:2013jaa}. Although early LHC data have seen this qualitative behavior~\cite{Khachatryan:2016qhq}, the STAR data has shown it explicitly for the first time, especially for the incoherent production.

\subsection{\label{subsec:interf}Interference}

\begin{figure}[thb]
\includegraphics[width=3.4in]{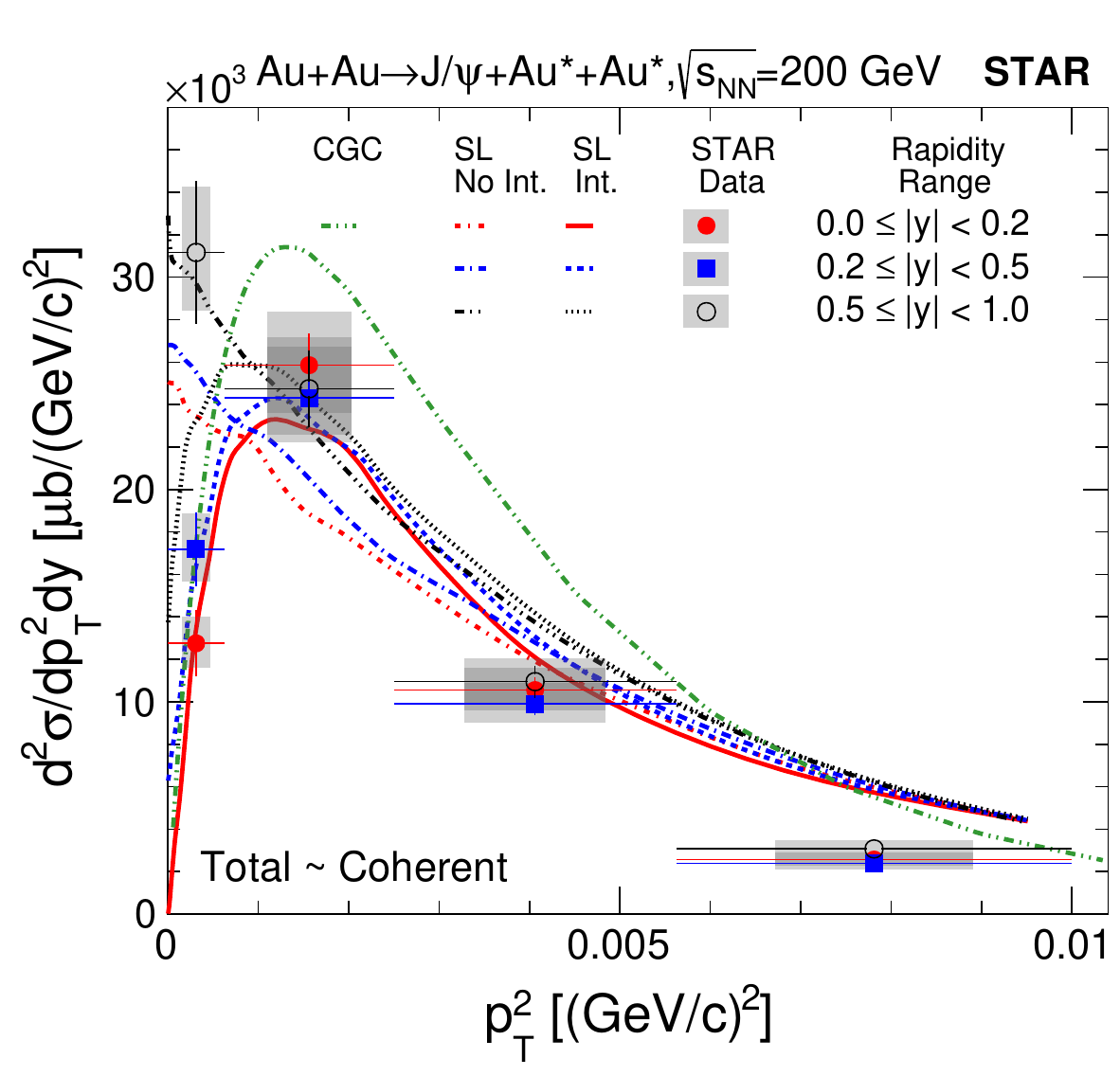}
  \caption{ \label{fig:res:figure_5_a} Differential cross section, $d^{2}\sigma/dp^{2}_{\mathrm{T}} dy$, as a function of \ptSquare at very low \ptSquare with different rapidity $|y|$ bins in \AuAu UPCs at \sNNrhic~GeV.
  All neutron categories are included.
  STARlight (SL) events and color glass condensate (CGC) calculations are compared with the data. There is a systematic uncertainty of 10\% from the integrated luminosity that is not shown. }
\end{figure}

Figure~\ref{fig:res:figure_5_a} shows the differential cross section of \jpsi photoproduction at very low \ptSquareNoSpace, for the three rapidity bins
and no selection of neutron category.
This region of \ptSquare is dominated by coherent photoproduction,
with contamination from incoherent processes of order 1\%.
In the lowest \ptSquare bin, the cross section at lowest rapidity is
suppressed more than 50\% relative to the highest rapidity bin,
with smaller suppression at intermediate rapidity.
In the higher \ptSquare bins cross sections are
approximately equal at all rapidities.

This is a result of quantum interference in symmetric
Au+Au UPCs due to the ambiguity of which nucleus is the hard photon source. It requires photons with opposite polarization in order to have such destructive interference~\cite{STAR:2019wlg,STAR:2022wfe,Li:2019yzy}. The effect of interference is shown quantitatively by the calculations of STARlight for both the cases of interference and no interference.
With no interference there is no suppression at lowest \ptSquareNoSpace.
STARlight with interference predicts the trend observed in the data,
suppression at lowest \ptSquare increasing as $y\rightarrow0$.
This interference effect has also been observed in
UPC $\rho^0$ photoproduction by the STAR collaboration~\cite{STAR:2008llz}.

The CGC model calculation, at $y=0$, also includes the effects
of interference.
The prediction for the cross section,
shown in Fig.~\ref{fig:res:figure_5_a},
describes the suppression at lowest rapidity and \ptSquare
observed in the data.

\subsection{\label{subsec:psi2s}$\psi(2s)$ cross section}

\begin{figure}[thb]
\includegraphics[width=3.4in]{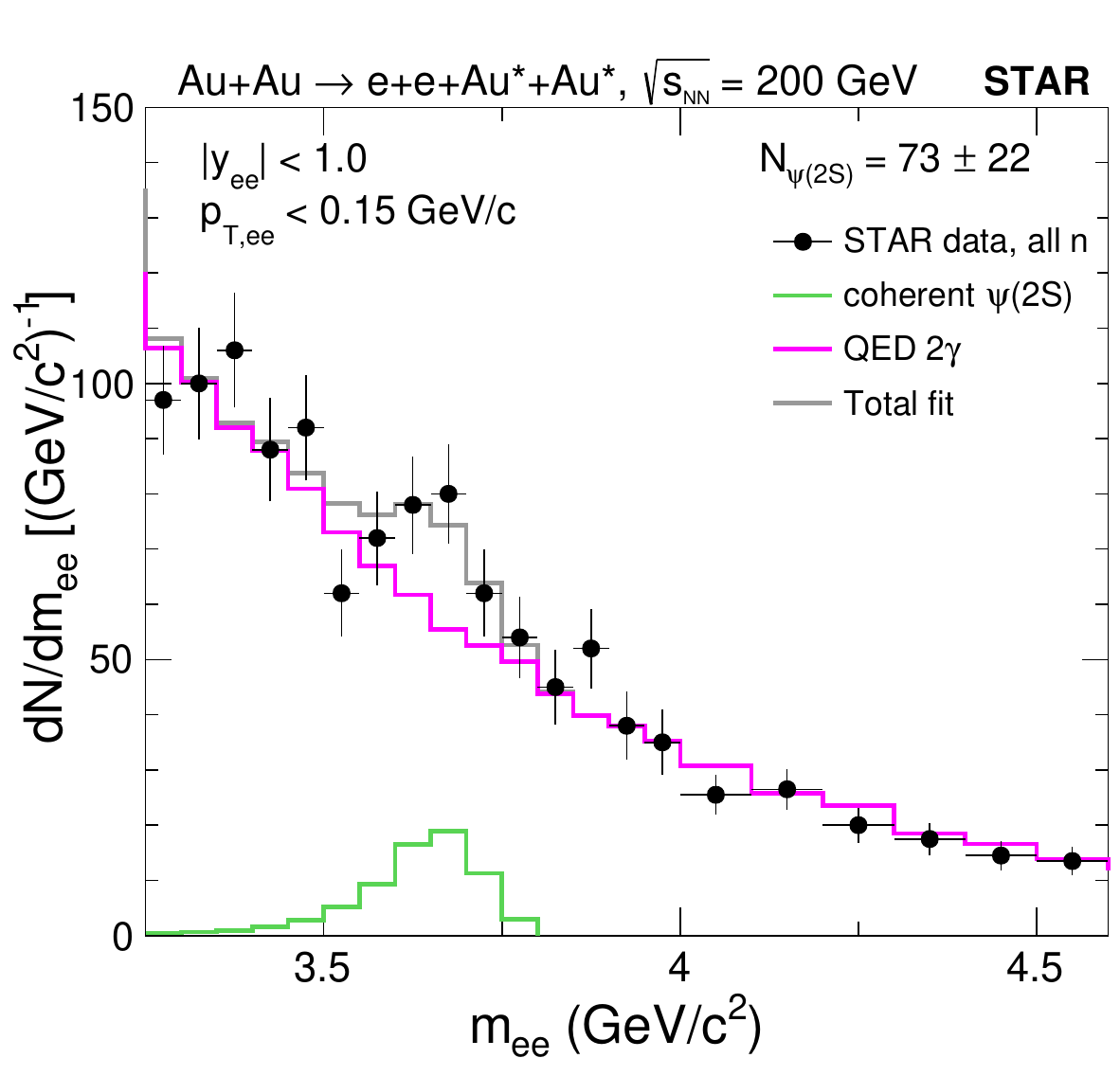}
  \caption{ \label{fig:res:figure_2}
  Invariant mass of the electron pair candidates from \AuAu UPCs at $\sqrt{s_{_\mathrm{NN}}}=200$ GeV. There is a systematic uncertainty of 10\% from the integrated luminosity that is not shown. }
\end{figure}

Figure~\ref{fig:res:figure_2} shows the template fit to
the raw $m_{\rm{ee}}$ distribution in the $\psi(2s)$ mass region.
The full rapidity range $|y|<1$ and all neutron categories are included.
The only processes contributing in this region are
QED $\gamma\gamma$ and coherent $\psi(2s)\rightarrow e^+ e^-$.
Their templates fit to the data and sum
are shown in the figure.
The number of $\psi(2s)$ events is $73 \pm 22$, where the
uncertainty is the statistical uncertainty from the fit.

Figure~\ref{fig:res:figure_8_b} shows the cross section
determined from this sample of events,
expressed as $d\sigma/dy$ in the range $0<|y|<1$.
The measured coherent \jpsi differential cross section
for $all~n$ is also shown for comparison.
The bottom panel shows the ratio of $\psi(2s)$ to \jpsi
cross sections. The predictions from STARlight are also shown, which exceeds the individual cross sections
by 20-30\% but correctly predicts the measured ratio.

\subsection{\label{subsec:NLO}Next-to-leading order perturbative QCD calculation}

\begin{figure}[thb]
\includegraphics[width=3.4in]{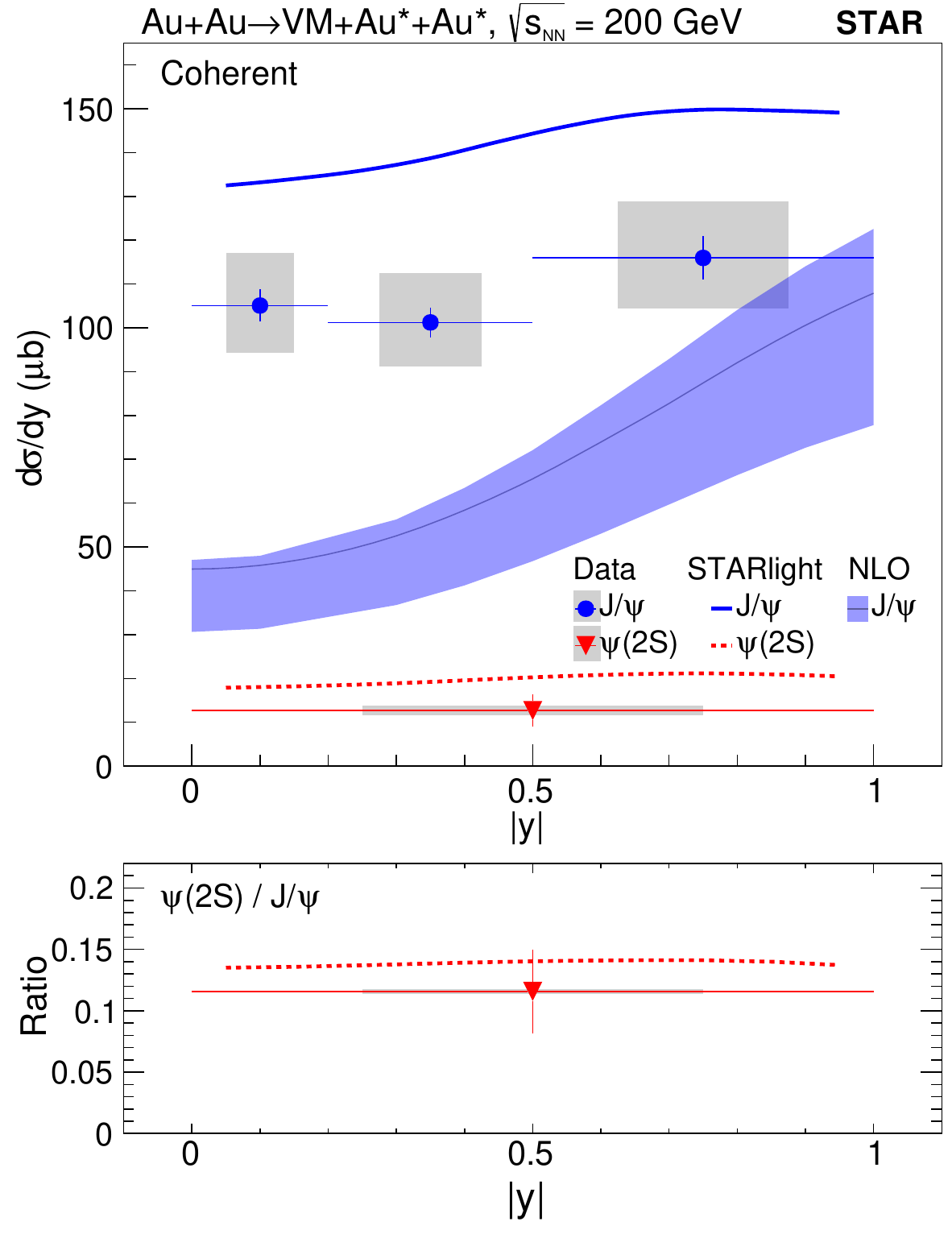}
  \caption{ \label{fig:res:figure_8_b} Differential cross section $d\sigma/dy$ for coherent \jpsi and $\psi(2s)$ photoproduction as a function of $|y|$ in \AuAu UPCs at \sNNrhic~GeV.
  The STARlight model~\cite{Klein:2016yzr}
  and the NLO pQCD calculations~\cite{Eskola:2022vaf,Eskola:2022vpi}
  are compared with the data. Ratio between $\psi(2s)$ and \jpsi is shown in the bottom panel. Statistical uncertainty is represented by the error bars, and the systematic uncertainty is denoted as boxes. There is a systematic uncertainty of 10\% from the integrated luminosity that is not shown.  }
\end{figure}

The colored band in 
Fig.~\ref{fig:res:figure_8_b}
shows the first NLO perturbative QCD calculation of the \jpsi photoproduction at RHIC energies. The input nuclear PDF (nPDF) is from EPPS21~\cite{Eskola:2021mjl}, where the current uncertainty coming from the nPDF on the \jpsi production cross section can be as large as 50\% to 160\%.
Consequently, this is not shown.
The uncertainty band shown on the figure is only based on the scale uncertainty. For details, see Refs~\cite{Eskola:2022vaf,Eskola:2022vpi}. This prediction has been found to be underestimated by more than a factor of 2 at mid-rapidity and 10-20\% at higher rapidity. This data will significantly constrain the nPDF at the NLO for both quarks and gluons.

\subsection{\label{subsec:twogamma}$\gamma\gamma\rightarrow e^{+}e^{-}$ cross sections}

\begin{figure}[t]
\includegraphics[width=3.4in]{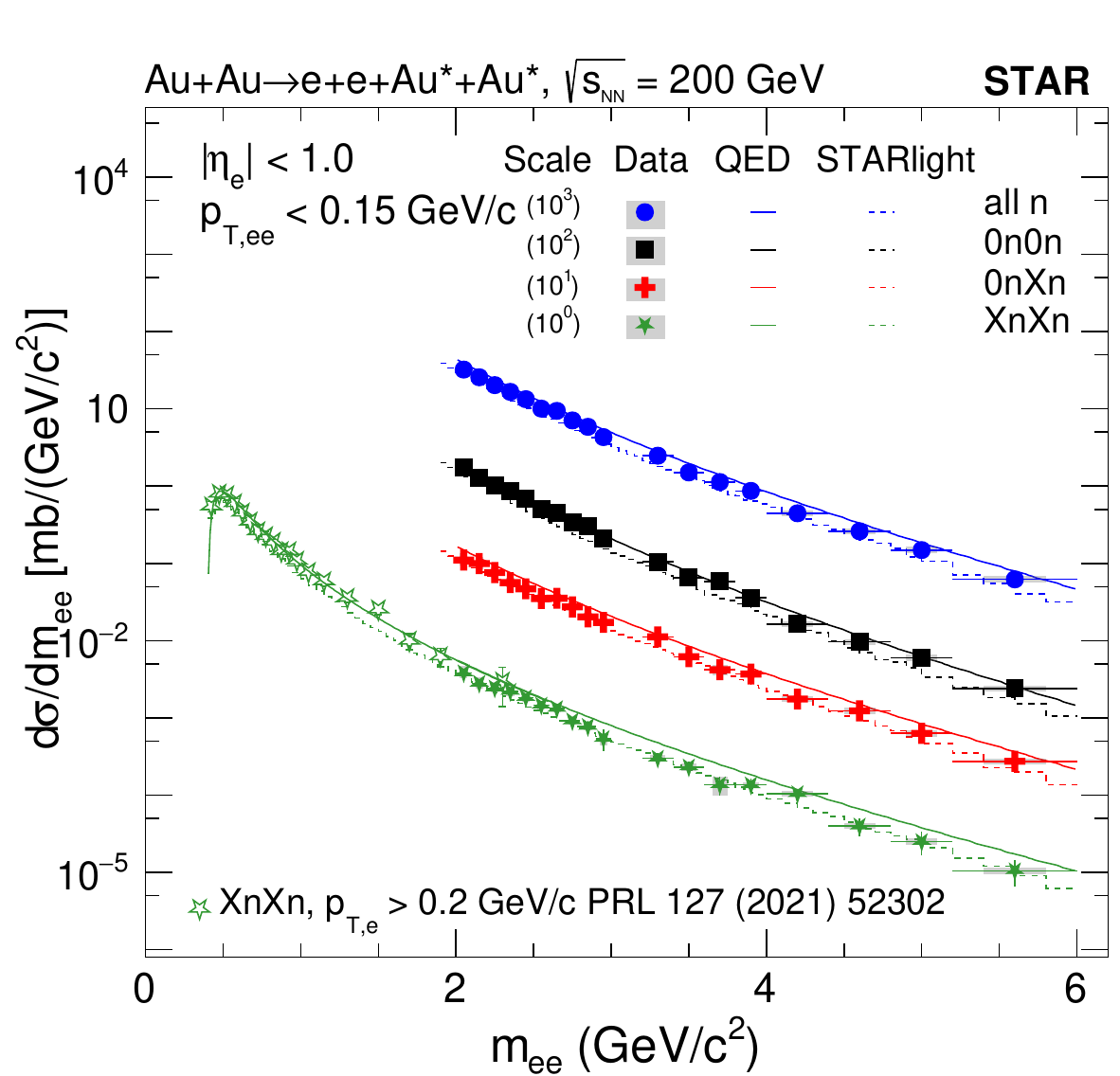}
\caption{ \label{fig:res:figure_11_a} Differential cross section of exclusive electron pair production as a function of electron pair ($ee$) invariant mass in \AuAu UPCs at \sNNrhic~GeV. The rapidity $y$ of the pair is within $\pm$ 1.0 unit, identical to that of the \jpsi particle. The
STARlight model~\cite{Klein:2016yzr} and QED theory calculations from Zha et al~\cite{Zha:2018tlq} are compared with data. There is a systematic uncertainty of 10\% from the integrated luminosity that is not shown. }
\end{figure}

\begin{figure}[t]
\includegraphics[width=3.4in]{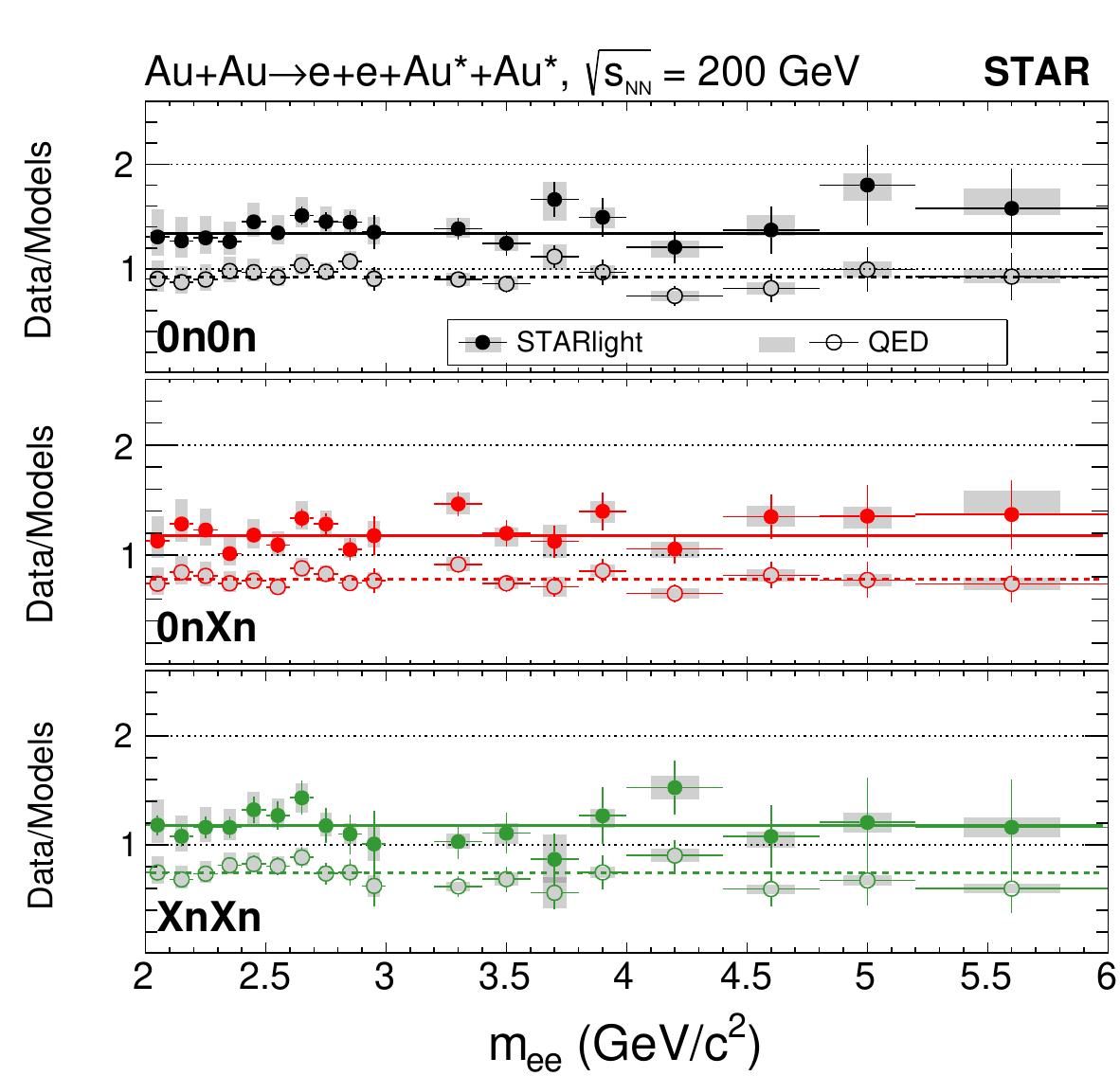}
\caption{ \label{fig:res:figure_11_b} The ratios between data and models are shown, where the models are
STARlight~\cite{Klein:2016yzr}
and QED theory calculations from Zha et al~\cite{Zha:2018tlq}.
Statistical uncertainty is represented by the error bars, and the systematic uncertainty is denoted as boxes. There is a systematic uncertainty of 10\% from the integrated luminosity that is not shown.}
\end{figure}

In Fig.~\ref{fig:res:figure_11_a}, the differential cross sections of $\gamma\gamma$ to $e^{+}e^{-}$ pairs, as a function of the pair mass $m_{\rm{ee}}$,
are shown for different neutron configurations. The data are compared with both STARlight and a QED calculation
performed by Zha et al~\cite{Zha:2018tlq,Zha:2018ywo}. The ratios between the data and these two predictions are shown in Fig.~\ref{fig:res:figure_11_b}.
The central value of the data are 10-20\% above the
STARlight prediction, and 10-20\%
below the QED calculation.
It should be noted that
STARlight does not include $e^{+}e^{-}$ pair
production inside the nucleus,
whereas the QED calculation does.
The magnitude of this effect has been
estimated to be
$\approx 10\%$~\cite{Klein:2016yzr},
partially accounting for the discrepancy seen in STARlight relative to the data.
The scale uncertainty from the luminosity measurement is 10\% (not shown), implying that both models are consistent with the data.

The consistency between data and models across different neutron categories
validates the photon fluxes used in the measurement of
$\gamma$+Au cross sections from Au+Au cross sections,
described in Sections~\ref{subsec:gamAuxsec} and~\ref{subsec:gammaAuxsecdat}
and shown in Fig.~\ref{fig:res:figure_9}.
The Au+Au cross sections are linear in the photon flux
$\Phi^{ntag}_{T, \gamma}$, as in Eq.~\ref{eqn:dsigdy_dsigdw}.
The QED $\gamma\gamma$ cross sections are as follows:
\be
\label{eqn:twophotoncrosssec}
\sigma_{\gamma\gamma} \propto
\int dk_1 dk_2 \Phi^{ntag}_{T, \gamma}(k_1) \Phi^{ntag}_{T, \gamma}(k_2)
|\mathcal{M}(k_1,k_2)|^2 \, .
\ee
\noindent Here $\mathcal{M}$, the QED matrix element for
$\gamma \gamma \rightarrow e^+ e^-$,
is well known. The $\gamma\gamma$ cross section is quadratic in the photon fluxes, while the Au$+$Au cross sections are linear in the flux. Therefore, the $\sigma_{\gamma\gamma}$ is a test of the photon fluxes.
The 10-20\% model and data discrepancy in
Fig.~\ref{fig:res:figure_11_b} implies that deviations in $\sigma_{\rm Au+Au}$ are half as large. Therefore, the fluxes used in the $\gamma$+Au cross sections measurement
are valid at the 5-10\% level. Other models, such as SuperChic~3~\cite{Harland-Lang:2018iur,Harland-Lang:2021ysd,Harland-Lang:2023ohq}, may be compared with the reported data in the future.

Figure~\ref{fig:res:figure_11_a} also shows the $\gamma\gamma\rightarrow e^{+}e^{-}$ cross section
for the $XnXn$ category from a previous STAR publication~\cite{STAR:2019wlg}.
That measurement at lower $m_{\rm{ee}}$
utilized different experimental techniques
than the present data, based on TOF as opposed to BEMC selections. The two data sets have excellent agreement in the overlap region near $m_{\rm{ee}} \approx 2$ GeV/c$^2$, providing further affirmation of the measurements. These results present the first $\gamma\gamma\rightarrow e^{+}e^{-}$ measurement up to an invariant mass of 6 $\rm GeV/c^{2}$.

\section{\label{sec:conclusion}Conclusion}
Exclusive $J/\psi$, \psiTwoS, and $e^{+}e^{-}$ pair photoproduction in Au$+$Au UPCs at \sNNrhic~GeV using the STAR detector are measured. For \jpsi photoproduction, both coherent and incoherent processes as a function of rapidity $y$, \ptSquareNoSpace, and different neutron configurations are presented. In particular, three different neutron configurations, $0n0n$, $0nXn$, and $XnXn$, are combined to resolve the photon energy ambiguity in UPCs, which leads to the total cross section of coherent \jpsi photoproduction in $\gamma +$Au collisions as a function of photon-nucleon center-of-mass energy. It is found that the coherent nuclear suppression factor at \wGammaN$=25.0$ GeV is $0.71\pm0.10$ when compared to the expectation of a free nucleon. This suppression supports the nuclear shadowing effect with the leading twist approximation. The Next-to-Leading order calculation of perturbative Quantum Chromodynamics on coherent \jpsi photoproduction is compared with this measurement. The description of the data is off by a factor of two at midrapidity based on nPDF EPPS21, which may indicate the large uncertainty on the nuclear parton distribution functions that this data can significantly constrain. The nuclear parton density at the top RHIC energy is at the region between large momentum quarks ($x_{\rm{parton}}>0.1$) and low momentum gluons ($x_{\rm{parton}}<0.001$), which is essential to the understanding of nuclear modification effects in this transition regime. Moreover, incoherent \jpsi photoproduction has been measured up to high \ptSquare of $2.2~\rm{[(GeV/c)^{2}]}$, and the incoherent suppression factor at \wGammaN$=25.0$ GeV is found to be $0.36\pm0.07$ relative to the free proton. Based on a hot-spot model with sub-nucleonic parton density fluctuation, the incoherent data indicate a similar level of fluctuation seen in the free proton as it is characterized by the shape of the \ptSquare distribution. However, direct comparisons between hot-spot models (Sartre and CGC) and data cannot be fully reconciled and further theory investigations are needed to draw a conclusion. Finally, the QED $\gamma\gamma \rightarrow e^{+}e^{-}$ has been measured up to an invariant mass of 6 $\rm GeV/c^{2}$ for different neutron emission classes, which constrains the modelling of neutron emission and photon flux. The data provide important constraints to the parton density and its fluctuations, and also provide an essential experimental baseline for such measurement at the upcoming Electron-Ion Collider.

\section*{Acknowledgements}
We thank the RHIC Operations Group and RCF at BNL, the NERSC Center at LBNL, and the Open Science Grid consortium for providing resources and support.  This work was supported in part by the Office of Nuclear Physics within the U.S. DOE Office of Science, the U.S. National Science Foundation, National Natural Science Foundation of China, Chinese Academy of Science, the Ministry of Science and Technology of China and the Chinese Ministry of Education, the Higher Education Sprout Project by Ministry of Education at NCKU, the National Research Foundation of Korea, Czech Science Foundation and Ministry of Education, Youth and Sports of the Czech Republic, Hungarian National Research, Development and Innovation Office, New National Excellency Programme of the Hungarian Ministry of Human Capacities, Department of Atomic Energy and Department of Science and Technology of the Government of India, the National Science Centre and WUT ID-UB of Poland, the Ministry of Science, Education and Sports of the Republic of Croatia, German Bundesministerium f\"ur Bildung, Wissenschaft, Forschung and Technologie (BMBF), Helmholtz Association, Ministry of Education, Culture, Sports, Science, and Technology (MEXT), Japan Society for the Promotion of Science (JSPS) and Agencia Nacional de Investigaci\'on y Desarrollo (ANID) of Chile.

\bibliography{reference}
\end{document}